\documentclass[aps,prd,reprint,preprintnumbers,superscriptaddress,nofootinbib]{revtex4-1}
\usepackage[all]{xy}
\usepackage{amsmath,amsthm,amssymb}
\usepackage[dvipsnames, usenames]{xcolor}
\usepackage{comment}
\usepackage{array}
\usepackage{bm}
\usepackage[normalem]{ulem} 
\allowdisplaybreaks[1]

\usepackage{hyperref}
\usepackage{tcolorbox}
\usepackage{mathrsfs}
\usepackage{float}
\usepackage{orcidlink}
\usepackage{tikz}
\usepackage{xspace}

\newcommand{\Eq}[1]{Eq.~\eqref{eq:#1}}
\newcommand{\Fig}[1]{Fig.~\ref{Fig:#1}}
\newcommand{\Figure}[1]{Figure~\ref{Fig:#1}}

\newcommand{\pin}{p_\infty}

\newcommand{\spec}{\textsc{SpEC}\xspace}
\newcommand{\ETK}{\textsc{ETK}}
\newcommand{\SEOBNR}{\texttt{SEOBNRv5}}
\newcommand{\SEOBPM}{\texttt{SEOB-PM}}
\newcommand{\SEOBfourPM}{\texttt{SEOB-4PM}}
\newcommand{\SEOBfivePM}{\texttt{SEOB-5PM(1SF)}}
\newcommand{\Dali}{\texttt{TEOBResumS-Dal\'{i}}}

\begin{document}

\preprint{HU-EP-25/25-RTG, QMUL-PH-25-15}

\title{Highly accurate simulations of asymmetric black-hole scattering and\\ cross validation of effective-one-body models}

\def\AEI{Max Planck Institute for Gravitational Physics (Albert Einstein Institute), D-14476 Potsdam, Germany}
\def\Humboldt{Institut für Physik und IRIS Adlershof, Humboldt-Universität zu Berlin, 10099 Berlin, Germany}
\def\QMUL{Centre for Theoretical Physics, Department of Physics and Astronomy, Queen Mary University of London,  London E1~4NS, United Kingdom}
\newcommand{\UIB}{{Departament de F\'isica, Universitat de les Illes Balears, IAC3 -- IEEC, Crta. Valldemossa km 7.5, E-07122 Palma, Spain}}

\author{Oliver Long \orcidlink{0000-0002-3897-9272}}
\email{oliver.long@aei.mpg.de}
\affiliation{\AEI}
\author{Harald P. Pfeiffer \orcidlink{0000-0001-9288-519X}}
\affiliation{\AEI}
\author{Alessandra Buonanno \orcidlink{0000-0002-5433-1409}}
\affiliation{\AEI}
\affiliation{Department of Physics, University of Maryland, College Park, MD 20742, USA}
\author{Gustav Uhre Jakobsen \orcidlink{0000-0001-9743-0442}}
\affiliation{\AEI}
\affiliation{\Humboldt}
\author{\\Gustav Mogull \orcidlink{0000-0003-3070-5717}}
\affiliation{\AEI}
\affiliation{\Humboldt}
\affiliation{\QMUL}
\author{Antoni Ramos-Buades \orcidlink{0000-0002-6874-7421}}
\affiliation{\UIB}
\author{Hannes R. Rüter\,\orcidlink{0000-0002-3442-5360}}
\affiliation{CENTRA, Departamento de F\'{\i}sica, Instituto Superior T\'ecnico -- IST, Universidade de Lisboa -- UL,
Avenida Rovisco Pais 1, 1049 Lisboa, Portugal}
\author{Lawrence E. Kidder \orcidlink{0000-0001-5392-7342}}
\affiliation{Cornell Center for Astrophysics and Planetary Science, Cornell University, Ithaca, New York 14853, USA}
\author{Mark A. Scheel \orcidlink{0000-0001-6656-9134}}
\affiliation{TAPIR 350-17, California Institute of Technology, 1200 E California Boulevard, Pasadena, California 91125, USA}

\date{\today}
\begin{abstract}
The study of unbound binary--black-hole encounters provides a gauge-invariant 
approach to exploring strong-field gravitational interactions in two-body systems, 
which can subsequently inform waveform models for bound orbits.
 In this work, 
we present 60 new highly accurate numerical relativity (NR) simulations of 
black-hole scattering, generated using the Spectral Einstein Code (\spec{}). Our 
simulations include 14 spin-aligned configurations, as well as 16 configurations with unequal masses, up to a mass ratio of 10.
We perform the first direct comparison of scattering angles computed 
using different NR codes, finding good agreement. We compare our NR scattering angle results to the post-Minkowskian (PM)-based effective-one-body (EOB) closed-form models \SEOBPM{}
 and $w_{\rm EOB}$, finding less than 5\% deviation except near the scatter-capture separatrix. Comparisons with the post-Newtonian-based EOB evolution models \SEOBNR{} and \Dali{} reveal that the former agrees within 8\% accuracy with non-spinning NR results across most parameter ranges, whereas the latter matches similarly at lower energies but diverges significantly at higher energies. Both evolution EOB
models exhibit increased deviations for spinning systems, predicting a notably different location of the capture separatrix compared to NR.
 Our key result is the first measurement of disparate scattering angles from NR simulations due to
asymmetric gravitational-wave emission. We compare these results to \SEOBPM{} models constructed to calculate the scattering angle of a single black hole in asymmetric systems.
\end{abstract}

\maketitle

\section{Introduction} 

\begin{figure}[htb]
  \centering
  \includegraphics[width=\linewidth,trim=10 9 6 6,clip=true]{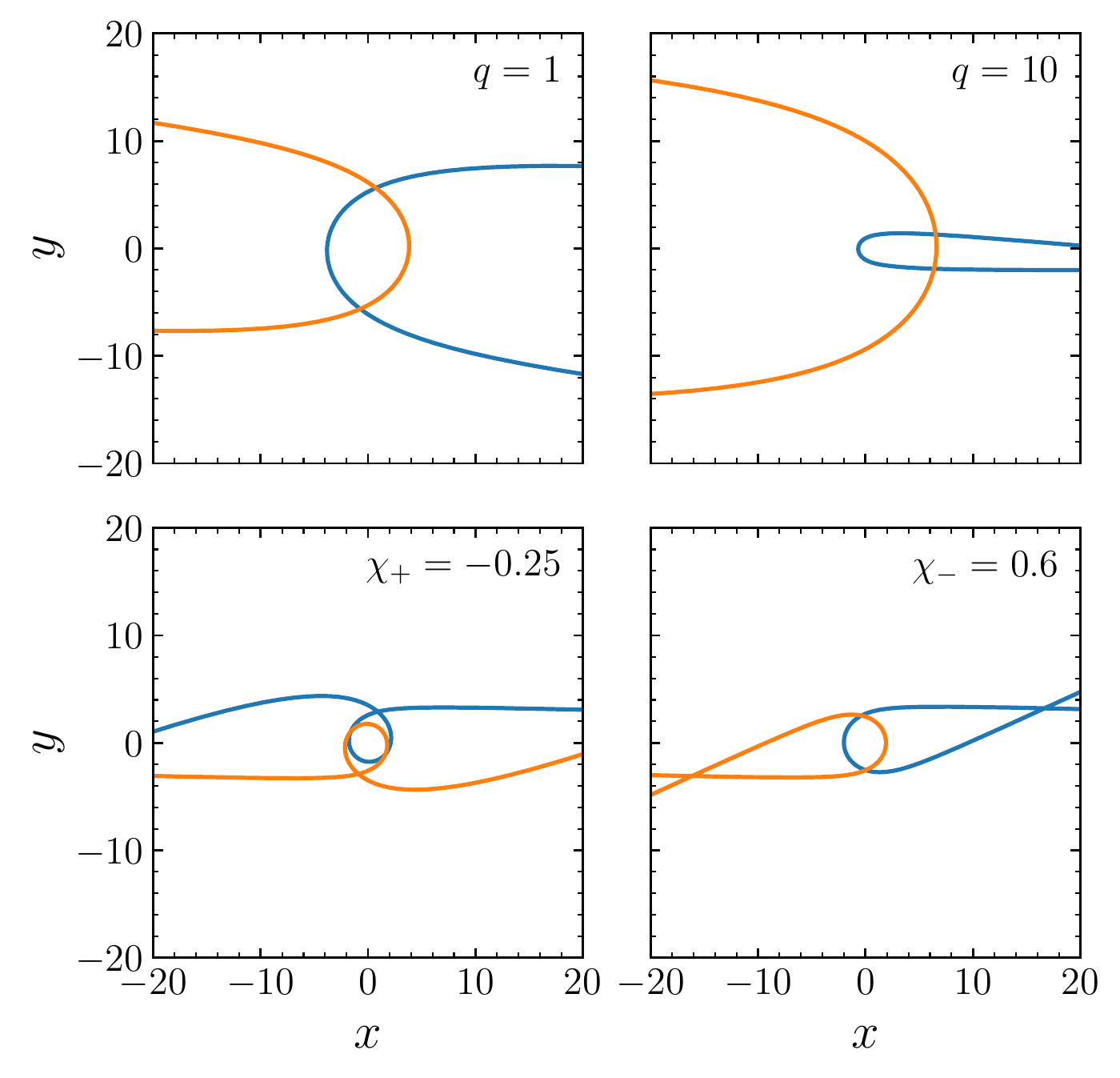}
  \caption{
    Coordinate trajectories of two unbound BHs for a variety of initial conditions.
  {\em Upper}: Non-spinning BHs with $\gamma=1.02$, $\ell = 4.8$, and mass ratios $q=1$ ({\em left}) and $q=10$ ({\em right}).
  {\em Lower}: Equal mass BHs with $\gamma=1.226$, $\ell = 5.18$, and equal parallel spins $\chi_+=-0.25$ ({\em left}) and equal magnitude, anti-parallel spins $\chi_-=0.6$ ({\em right}). See Eqs.~(\ref{eq:qDef}), (\ref{eq:gammaDef}), (\ref{eq:ellDef}), and (\ref{eq:chipmDef}) for definitions of $q$, $\gamma$, $\ell$, and $\chi_\pm$, respectively.}
  \label{Fig:Orbits}
  \end{figure}

  Observations of gravitational waves (GWs) from binary--black-hole (BBH) coalescence by the LIGO-Virgo-KAGRA Collaboration have opened a 
  new era in astrophysics and fundamental physics~\cite{LIGOScientific:2016aoc,LIGOScientific:2017vwq, LIGOScientific:2018mvr, LIGOScientific:2020stg, LIGOScientific:2020ibl, LIGOScientific:2021usb, KAGRA:2021vkt}. These detections enable us to probe the strong-field regime of General Relativity (GR) 
  with unprecedented precision. Unlike traditional electromagnetic astronomy, 
  GW observations require prior knowledge of the waveform in order 
  to extract the signal from noise-dominated data.

  We can simulate strong-field binary coalescence using numerical relativity (NR)~\cite{Pretorius:2005gq, Campanelli:2005dd, Baker:2005vv}.
  Here, the full Einstein field equations, ten highly-nonlinear partial differential 
  equations, are discretized and solved directly on supercomputers. These simulations provide a detailed, fully general-relativistic 
  description of phenomena and capture all of the physics of the system, but are  
  computationally expensive and limited in parameter space.
  
  In order to obtain the speed and accuracy required for data analysis, we need to develop
  fast and accurate waveform models~\cite{Abac:2025saz,LISAConsortiumWaveformWorkingGroup:2023arg, lvk_obs_whitepaper2024}. Effective-one-body (EOB)~\cite{Buonanno:1998gg, Buonanno:2000ef, Damour:2000we, Damour:2001tu, Buonanno:2005xu, Buonanno:2006ui}
  is one of the most successful approaches to modeling the dynamics and
  gravitational waveforms of binary systems. The EOB formalism is based on the
  idea of mapping the dynamics of a binary system to that of a single 
  effective particle moving in a curved spacetime. This is obtained by
  deforming a Schwarzschild (or Kerr) Hamiltonian with terms from perturbative 
  expansions (post-Newtonian~\cite{Ramos-Buades:2023ehm,Gamba:2021ydi}, post-Minkowskian~\cite{Buonanno:2024byg, Damour:2025uka}, self-force~\cite{vandeMeent:2023ols,Leather:2025nhu}, etc.), and via calibration to NR data~\cite{Pompili:2023tna, Nagar:2018zoe}.
  The resummed EOB Hamiltonian is evolved via Hamilton's equations,
  with additional terms to account for the radiation reaction.
  Current state-of-the-art ``evolution"\footnote{Here, by evolution models we mean models obtained by solving Hamilton's equations in time domain.} EOB models fall into two families: \texttt{SEOBNR}~\cite{Pompili:2023tna, Ramos-Buades:2023ehm, vandeMeent:2023ols, Khalil:2023kep, Gamboa:2024hli,Buonanno:2024byg,Leather:2025nhu} and \texttt{TEOBResumS}~\cite{Nagar:2018zoe, Nagar:2019wds, Nagar:2020pcj, Gamba:2021ydi, Nagar:2023zxh, Albanesi:2025txj}.
  
  As detector sensitivity improves, so 
  must the accuracy of our waveform models to avoid systematic 
  biases in parameter recovery due to modeling error, which 
  would diminish the science return from the improved sensitivity~\cite{Dhani:2024jja, Purrer:2019jcp}. 
  Additionally, as detectors become sensitive to a broader range of 
  physical properties (larger mass asymmetry, eccentricity, spin 
  configurations), waveform models need to be updated to cover the relevant parameter space. Both the recent science case for the Einstein Telescope~\cite{Abac:2025saz}
  and the White Paper commissioned by the LISA Consortium~\cite{LISAConsortiumWaveformWorkingGroup:2023arg}
  call for accelerated progress on BBH waveforms and highlight
  the need for fresh ideas.

  One promising approach is based on the study of BH scattering encounters.
  The idea here is to use information of the high-energy
  unbound interactions to inform models of bound systems~\cite{Damour:2016gwp,Damour:2017zjx}. An advantage 
  of scattering encounters is that we can define the initial conditions and certain observables in terms of
  asymptotic quantities defined at past/future infinity. This allows us to avoid the 
  local gauge ambiguities that plague comparisons of bound systems, 
  where the initial conditions are defined at a finite separation.
  The scattering angle is a particularly useful asymptotic observable as it is 
  directly related to the deflection of the BHs due to their 
  gravitational interactions. As illustrated in \Figure{Orbits}, variations in the initial conditions can significantly alter the black hole trajectories and, in turn, the resulting scattering angle.

  Inspired by these ideas, NR codes that were 
  originally designed to model the late inspiral and merger of 
  compact binaries have been adapted to study scattering encounters.
   Unbound NR
  simulations have been used to extract the scattering angle in BBH~\cite{Damour:2014afa,Damour:2022ybd,Hopper:2022rwo,Rettegno:2023ghr,Albanesi:2024xus,Swain:2024ngs} and binary-neutron-star~\cite{Fontbute:2025vdv} systems, explore the precise location of the 
  scatter-capture separatrix~\cite{Pretorius:2007jn,Sperhake:2008ga,Witek:2010xi,
  Sperhake:2010uv,Sperhake:2012me,Sperhake:2015siy}, test the boundary-to-bound (B2B) correspondence~\cite{Kankani:2024may}, extract induced 
  spin~\cite{Nelson:2019czq,Jaraba:2021ces,Rodriguez-Monteverde:2024tnt} and 
  quasinormal mode excitations~\cite{Bae:2023sww} from close hyperbolic 
  encounters, as well as the construction of a surrogate
  model for the Weyl scalar $\psi_4$ from scattering orbits~\cite{Fontbute:2024amb}.

  Scattering observables (like the angle) are conveniently described using a perturbative
  post-Minkowskian (PM) expansion~\cite{Westpfahl:1985tsl},
  wherein one expands in powers of the inverse separation between the two BHs ---
  or, equivalently, powers of Newton's constant $G$.
  The PM regime is considered valid when this separation is large compared with
  the BHs intrinsic size, characterized by the radius of the horizon.
  Thus, the BHs may be interpreted as point masses.
  Inspired by this pointlike view of BHs,
  recent calculations of scattering observables for BBH encounters have borrowed heavily from collider physics~\cite{Goldberger:2004jt,Bjerrum-Bohr:2022blt,Kosower:2022yvp,Buonanno:2022pgc},
  particularly techniques for performing multi-loop Feynman integrals ---
  including integration-by-parts identities (IBPs)~\cite{Laporta:2000dsw,Lee:2012cn,Smirnov:2019qkx,Klappert:2020nbg} and differential equations (DEs)~\cite{Kotikov:1990kg,Gehrmann:1999as,Henn:2013pwa}.

Recent state-of-the art PM results include
  the complete 5PM(1SF) scattering angle~\cite{Driesse:2024xad,Driesse:2024feo},
  produced using the Worldline Quantum Field Theory formalism~\cite{Mogull:2020sak,Jakobsen:2022psy}.
  Here first self-force order (1SF) indicates that this result is valid
  only up to terms linear in the symmetric mass ratio $\nu$, defined in \Eq{nuDef}.
  In this paper we also make use of 4PM non-spinning~\cite{Dlapa:2022lmu,Dlapa:2023hsl,Damgaard:2023ttc}
  and 5PM spinning results~\cite{Jakobsen:2022fcj,Jakobsen:2022zsx,FebresCordero:2022jts,Jakobsen:2023ndj,Jakobsen:2023hig},
  which have been obtained using both worldlines and scattering amplitudes.
  Our PM counting convention used here is a physical one,
  and includes factors of $G$ arising from higher powers of spin.
  Other recent work includes a 7PM scattering angle at quartic order in spin on one BH~\cite{Akpinar:2025bkt},
  and an extraction of the local-in-time part of the scattering angle relevant also
  for describing bound orbits~\cite{Bini:2024tft,Dlapa:2024cje,Dlapa:2025biy}.

  While highly successful, the PM expansion becomes inaccurate as the separation 
  between the bodies decreases and strong-field effects from the large curvature can no longer be
  neglected. Resummation techniques have incorporated information about the divergence of
  the scatter-capture separatrix to provide more faithful models across all separations~\cite{Damour:2022ybd,Long:2024ltn}.
  EOB resummation shines in the context of a PM expansion because,
  as we shall review in Section~\ref{sec:EOB},
  an EOB model may be matched order-by-order to gauge-invariant scattering observables~\cite{Damour:2016gwp,Damour:2017zjx}.
  The resulting ``closed-form'' EOB models can be used to make direct predictions for the scattering
  angle in the non-perturbative regime,
  without the need to fully evolve the system through Hamilton's equations.
  Instead, the angle can be extracted by numerically evaluating a suitable integral.

  The structure of the paper is as follows. In Sec.\ \ref{sec:NRSimulations} 
  we summarize the Spectral Einstein Code (\spec{}) and the extensions we 
  have made to the code to enable scattering encounters, detail our procedure to extract the 
  scattering angle, and perform a direct comparison between scattering angle 
  results from different NR codes. Section \ref{sec:EOB} reviews the EOB 
  formalism and introduces the models used to calculate the scattering angle, 
  namely the evolution model \SEOBNR{}, and the closed-form models 
  \SEOBPM{} and $w_{\rm EOB}$. We compare the scattering
  angle results from \spec{} and the EOB models in Sec.\ 
  \ref{sec:CompareToEOB} for a variety of initial conditions including 
  aligned spins, and unequal masses. We conclude in Sec.\ 
  \ref{sec:Conclusion} with a summary of our findings and provide an outlook.

\subsection*{Notation}

Throughout this work, we use natural geometrised units with $G=c=1$. Our setup consists of two BHs with masses $m_1$ and $m_2$ with $m_1 \geq m_2$.
We denote the total mass, mass ratio, symmetric and asymmetric mass ratios by
\begin{subequations}
\begin{align}
M &= m_1 + m_2, \\
q &= \frac{m_1}{m_2}, \label{eq:qDef}\\
\nu &= \frac{m_1 m_2}{M^2} = \frac{\mu}{M},\label{eq:nuDef}\\
\delta &= \frac{m_1-m_2}{M},
\end{align}
\end{subequations}
where $\mu$ represents the reduced mass.
We utilize a rescaled version of the Arnowitt–Deser–Misner (ADM) energy, $E_{\rm ADM}$, defined by
\begin{equation}
\Gamma = \frac{E_{\rm ADM}}{M} = \sqrt{1+2\nu(\gamma-1)},
\end{equation}
where $\gamma$ is the relative Lorentz factor given by
\begin{equation}\label{eq:gammaDef}
\gamma = \frac{\Gamma^2M^2 -m_1^2 - m_2^2}{2m_1m_2}.
\end{equation}

All BHs considered in our work have their spin vector parallel to the orbital angular-momentum vector, and are characterized by the projection of the spin onto the angular momentum direction, denoted by $S_{i}, i=1,2$. We obtain the orbital angular momentum $L$ from the ADM angular momentum
\begin{equation}\label{eq:TotalAngularMomentum}
{J}_{\rm ADM} = {L} + {S}_1 + {S}_2,
\end{equation}
and introduce a rescaled orbital angular momentum
\begin{equation}
\ell = \frac{L}{m_1 m_2}. \label{eq:ellDef}
\end{equation}
In terms of these quantities, the impact parameter is given by
\begin{equation}\label{eq:bDef}
b = \frac{\ell \: \Gamma}{\sqrt{\gamma^2-1}}. 
\end{equation}
We define the dimensionless spins as $\chi_i = a_i/m_i = S_i/m_i^2$ which have a range $-1 \leq \chi_i \leq 1$.
  For spins aligned with the orbital angular momentum, it is useful to define the following combinations
\begin{equation}
  \chi_\pm = \frac{\chi_2 \pm \chi_1}{2}. \label{eq:chipmDef}
\end{equation}
Finally, we note that the ADM angular momentum defined through 
Eq.~\eqref{eq:TotalAngularMomentum} may be identified with the 
\textit{canonical} angular momentum, which is different from the 
\textit{covariant} angular momentum commonly used in post-Minkowskian literature~\cite{Buonanno:2024vkx}.

\section{Numerical Relativity simulations}
\label{sec:NRSimulations}

The work presented here uses the Spectral Einstein Code
(\spec{})~\cite{SpECwebsite}, a multi-domain spectral code for
the general relativistic initial-value problem and evolution problem.
The employed numerical techniques are summarized in
Refs.~\cite{Mroue:2013xna,Boyle:2019kee,Scheel:2025jct}.  In particular, \spec{}
evolves a first-order representation of the generalized harmonic
evolution system~\cite{Lindblom:2005qh} using a multi-domain spectral
method~\cite{Kidder:1999fv, Scheel:2008rj, Szilagyi:2009qz,
  Hemberger:2012jz}.  At the outer boundary \spec{} employs
  constraint-preserving  
boundary conditions~\cite{Lindblom:2005qh, Rinne:2006vv, Rinne:2007ui}
with approximately no incoming gravitational radiation,
whereas BH excision is used inside the apparent
horizons~\cite{Scheel:2008rj, Szilagyi:2009qz, Hemberger:2012jz,
  Ossokine:2013zga}.  The elliptic solver of
\spec{}~\cite{Pfeiffer:2002wt,Ossokine:2015yla} utilizes the
extended conformal-thin sandwich (XCTS) approach
\cite{York:1998hy,Pfeiffer:2002iy,Cook:2004kt}.

To enable accurate and efficient simulations of hyperbolic
  encounters, we extend and improve \spec{}'s initial data
  routines, as well as evolution algorithm, as described in
  Secs.~\ref{sec:InitialDataImprovements}
  and~\ref{sec:EvolutionImprovements}.  We then perform a variety of
NR simulations, as summarized in
Sec.~\ref{sec:ParameterSpaceCoverage}.
Section~\ref{sec:ExtactScatteringAngle} describes our techniques of
extracting the scattering angle with error estimates, which we compare to results in the literature 
in Sec.~\ref{sec:CompareToOtherNR}.

\subsection{Improvements: Initial data}
\label{sec:InitialDataImprovements}

  In order to allow sufficient flexibility in the constructed
  BBH initial data sets, \spec{}'s initial-data solver~\cite{Cook:2004kt,Ossokine:2015yla} utilizes
  several different sets of input parameters, that control different
  properties of the constructed initial-data sets:
  \begin{enumerate}
    \item Parameters that
  allow control over the masses, and spins of the BHs to be
  constructed~\cite{Cook:2004kt,Caudill:2006hw}.
\item Parameters that determine the trajectories of the binary~\cite{Pfeiffer:2007yz,Buchman:2012dw,Ossokine:2015yla}, notably the initial separation $D_0$,  orbital velocity
  $\Omega_0$, and the initial expansion
  $\dot{a}_0 = \dot{D}/D_0$.
  \item Parameters that control the position and velocity of the
    center of mass, to construct initial data in the center-of-mass
    frame~\cite{Ossokine:2015yla}.
  \item Finally, there are choices of various 3-dimensional fields
    (conformal metric, trace of the extrinsic curvature and their
    time-derivatives), as well as boundary conditions at the BH
    horizons and spatial infinity.  These choices impact the physics
    of the initial-data sets, for instance the maximum reachable BH
    spin~\cite{Lovelace:2008tw} or the strength and properties of an initial relaxation
    phase, where the geometry of the initial data relaxes to an
    equilibrium state, emitting gravitational radiation in the process (``junk radiation''~\cite{Lovelace:2008hd}).
  \end{enumerate}
  The present work required improvements and changes to several
  of these sectors.

  The initial separation $D_0$ is at least an order
  of magnitude larger for hyperbolic encounters than for
  quasi-circular BBH inspirals.  To achieve accurate solutions, the
  inner spheres (Sphere A and B in the notation of Ref.~\cite{Ossokine:2015yla})
  may now be split into multiple radial shells,
  such that each shell has a ratio of outer to inner radius of $\sim
  8$.

  The velocities of the BHs can also be larger than for quasi-circular
  configurations, in particular for cases with high initial energy and
  for the smaller BH in unequal-mass binaries.  The geometric
  relations between various elements of the domain-decomposition were
  adjusted to handle the correspondingly more extreme 
  Lorentz length contraction.

  As stated under 2.~above, the initial data code requires
  $\Omega_0$ and $\dot a_0$ as part of its input parameters.  However,
  for our scientific goals, it is much more convenient to be able to specify
  ADM energy and ADM angular momentum of the initial data sets.  We
  included new root-finding functionality that adjusts $\Omega_0$ and
  $\dot{a}_0$ such that the final initial-data set achieves
  user-specified $E_{\rm ADM}$ and $J_{\rm ADM}$. Details of this routine
  are presented in Secs.\ II D and E of Ref.~\cite{Mendes:2025gov}.

  We furthermore updated the preconditioner used in the elliptic solver,
  where we now use an additive Schwarz method with overlap
  1~\cite{SBjorstadG1996}, and with a full LU decomposition of each
  sub-block.  This could be accomplished by modifying options
  passed into the PETSc
  toolkit~\cite{petsc-web-page,petsc-user-ref,petsc-efficient}, which
  we utilize for the actual solution of the ensuing large non-linear
  systems of equations~\cite{Pfeiffer:2002wt}.  These changes increase
  the robustness of convergence of the PETSc linear solvers (via PETSc's
  sparse linear solvers), reduce the overall computational cost and
  also lead to a more accurate determination of $E_{\rm ADM}$ and
  $J_{\rm ADM}$. 

  Finally, all results reported here utilize conformal
  quantities of the ``Superposed Harmonic''
  form~\cite{Varma:2018sqd}, instead of ``Superposed
  Kerr-Schild''~\cite{Lovelace:2008tw} which has been more commonly
  used by quasi-circular \spec{} simulations.  While
  ``Superposed Harmonic'' initial data cannot handle BHs with
  nearly extremal spins, it leads to reduced initial transients and
  ``junk radiation''.  This is a particularly important advantage for hyperbolic
  encounters because one is limited in how long one can wait for
  initial transients to decay away (see
  Sec.~\ref{sec:EvolutionImprovements}).

\subsection{Improvements: \spec{} evolution}
\label{sec:EvolutionImprovements}

\subsubsection{Changes to the adaptive mesh refinement}
\label{sec:AMR}

\spec{} uses adaptive mesh refinement (AMR), which allows the numerical domain to automatically split or combine subdomains to ensure that the simulation has sufficient accuracy throughout the domain while keeping computational costs reasonable.
Simulations of hyperbolic encounters require several changes to the AMR algorithm of \spec{}, which previously was tuned to quasi-circular inspirals.

At the beginning of a simulation the BHs 
emit spurious high-frequency
radiation, which has been coined ``junk radiation''.
In typical inspiral simulations the existence of these
initial transients is tolerated and the properties of the system
are extracted when the system has relaxed and high frequency
transients in the initial data have decayed~\cite{Pretto:2024dvx}.
In the hyperbolic case,
we must start the simulation with a very large initial separation
of the BHs, otherwise initial transients would distort the
dynamics in the critical phase around periastron.
The large initial separation also has the benefit that 
the approximation of superposed
BH metrics used in the initial data construction is physically
better justified, thus reducing the amplitude of the
initial transients. In Appendix~\ref{App:Junk}, we investigate 
 the effect of the junk radiation 
and the consequences on the final scattering angle.

\begin{figure}
    \centering
    \includegraphics[width=\linewidth]{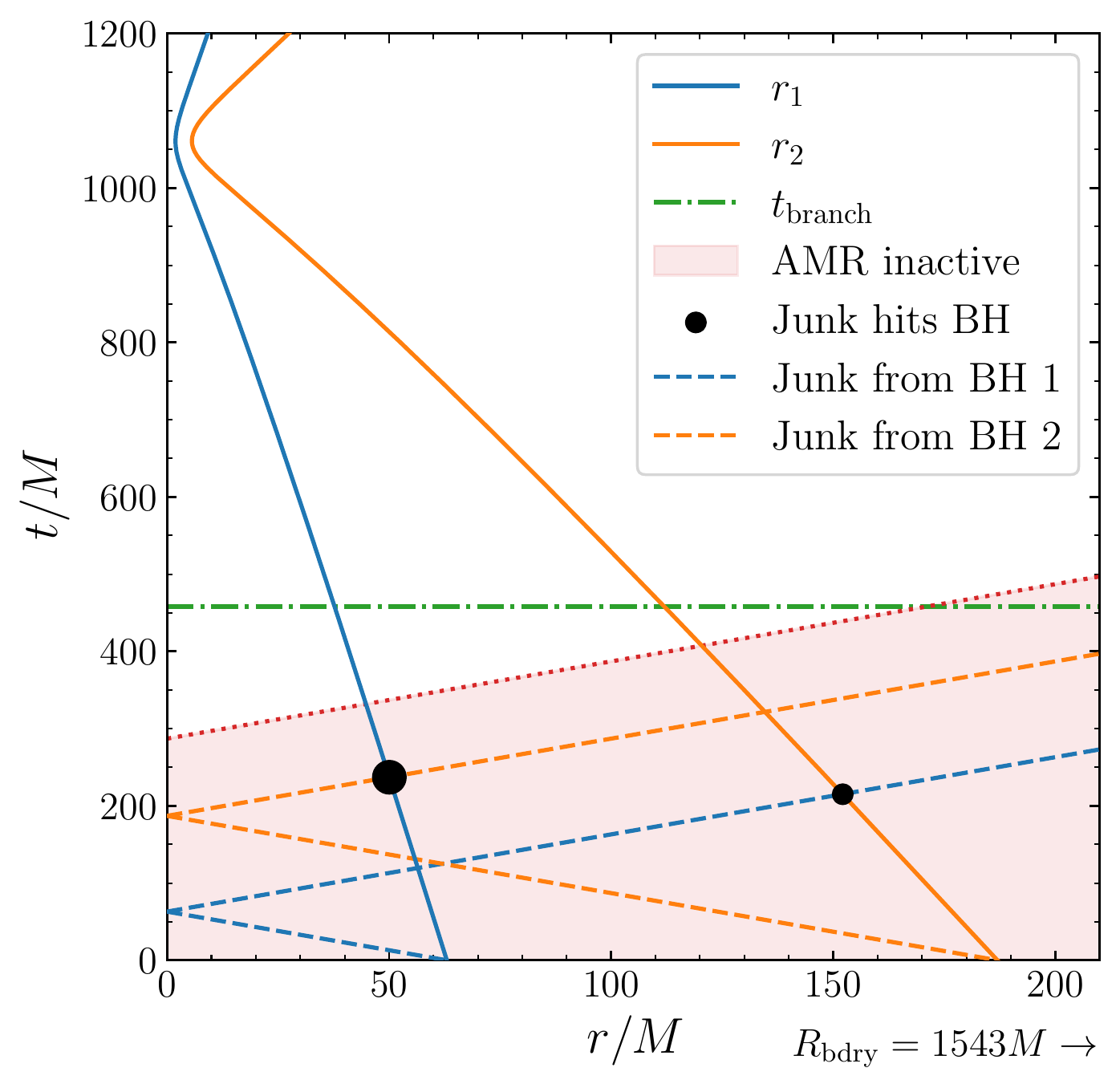}
    \caption{Spacetime diagram of a $q=3$ \spec{} scattering simulation. 
    The solid blue and orange lines show the trajectories of the 
    large and small BHs, respectively. The dashed lines 
    represent the junk radiation emanating from the BH of the matching color, 
    and the black discs show when the junk from one BH reaches the other.
    In the red-shaded region, encompassing the propagation of junk radiation, AMR is inactive and the red 
    dotted line shows when AMR is turned on.  The line $t_{\rm branch}$ marks the start of the lower resolution runs.
    We show the value of the outer boundary of the computational domain, $R_{\rm bdry}$ for reference.
    }
    \label{Fig:SpacetimeDiagram}
    \end{figure}

Another complication of junk radiation is that AMR would request an
excessive amount of resolution to resolve its high-frequency
components.  This would increase the computational cost of the
  simulations significantly, leading in extreme cases even to an
  exhaustion of computer memory.  To
  avoid such problems, the standard ``quasi-circular'' \spec{}
  activates AMR only after 1.2 light-crossing times to the outer
  boundary, which usually translates to $t\gtrsim 1000M$. Before that
  time, AMR in the outer spherical shells and the cylindrical domains
  between the individual BHs is disabled, and AMR in the inner
  spherical shells surrounding each BH is reduced in frequency of
  their update.  Because hyperbolic encounters proceed to the
  physically most relevant phase near periastron on much faster
  time-scales, the activation of AMR must be controlled more
  finely.  Figure~\ref{Fig:SpacetimeDiagram} illustrates the AMR
  algorithm used. At each spatial position, we estimate the 
  propagation of the main pulse of junk radiation (the dashed lines in
  Fig.~\ref{Fig:SpacetimeDiagram}), and activate AMR about $100M$
  thereafter, indicated by the dotted red line.
  In particular, this implies that the strong-field
  region has already active AMR, while the junk radiation still
  propagates toward the outer boundary.
  Because each computational element gets
  remapped by time-dependent maps, this requires to
  constantly monitor the distance of each grid-point in physical coordinates.
We continue to allow AMR with reduced cadence near each BH starting at $t=0$ in order to ensure well-resolved BHs with minimal impact of numerical truncation error.

While we throttle adaptive mesh refinement in the early phase of the simulation,
at later times, AMR must act faster than for quasi-circular inspirals,
in order to resolve the fast dynamics of the system.
To achieve that we decrease the minimal time span 
between two refinements of a subdomain element by 
a factor 100 as compared to standard \spec{} evolutions, 
concretely subdomain elements to refine every $0.02 m_2$.

Finally, near periastron passage, significant refinement is needed
close to the BHs. 
Previously, AMR was splitting subdomains, that are adjacent to the excision surface, only in the radial 
direction, but it can now also perform h-refinement to subdivide in the
angular directions around each BH.

\subsubsection{Methodology for convergence tests}

The fact that we choose to not resolve the high-frequency junk radiation
has consequences for the physical parameters of the evolved system.
If we start all resolutions of a convergence test at the beginning of the simulation,
each resolution will be differently affected by the unresolved part of the evolution,
leading to non-convergent differences after junk radiation has left. In order to mitigate this
effect, we follow the strategy discussed in Appendix~B of Ref.~\cite{VarFieSch19}. 
We only evolve the highest resolution from $t=0$ until time $t_{\rm branch}$.  We then consider this high-resolution data at $t=t_{\rm branch}$ as initial 
data for the lower resolutions, which we start at $t=t_{\rm branch}$ by downscaling the high-resolution run.
This ensures that each resolution is evolving the same physical system.
For the simulations we present here, the branching occurs at a time 
after the BHs have relaxed following the passage of the junk radiation (see \Fig{SpacetimeDiagram}). 
This choice ensures that the branching occurs before the close encounter,
but is too short for junk radiation to leave the domain. However, the most
severe impact of junk radiation is avoided.

\subsection{NR simulations}
\label{sec:ParameterSpaceCoverage}

    The parameter space of scattering simulations can be unambiguously 
    defined in terms of quantities defined at spatial infinity 
    $\Gamma$, $J_{\rm ADM}/(m_1 m_2)$, 
    as well as intrinsic BH properties ($q$, $\chi_i$). 
    The overall scale invariance of vacuum GR is represented by suitable 
    rescalings to make all these parameters dimensionless.
    For the unequal-mass case, it is more convenient to rescale the asymptotic
    quantities to those that tend to the geodesic limit as $\nu\rightarrow 0$,
    such as $\gamma$ and $\ell$ 
    defined in Eqs.~\eqref{eq:gammaDef} and \eqref{eq:ellDef} respectively. The \spec{} simulations start with the 
    BHs in the center-of-mass frame
        at Cartesian coordinates 
$\vec c_A \approx (D_0/(1+q), 0, 0)$ and $\vec c_B \approx (-D_0 \, q/(1+q), 0, 0)$ (GR corrections cause small deviations relative to these Newtonian values, see Ref.~\cite{Ossokine:2015yla}).  We end the simulations when the BHs reach a separation of $D = 1.4 D_0 $ on the 
    outgoing leg. For all the \spec{} simulations presented here, we use $D_0=250M$. 

\begin{figure}
\centering
\includegraphics[width=\linewidth,trim=12 10 6 6,clip=true]{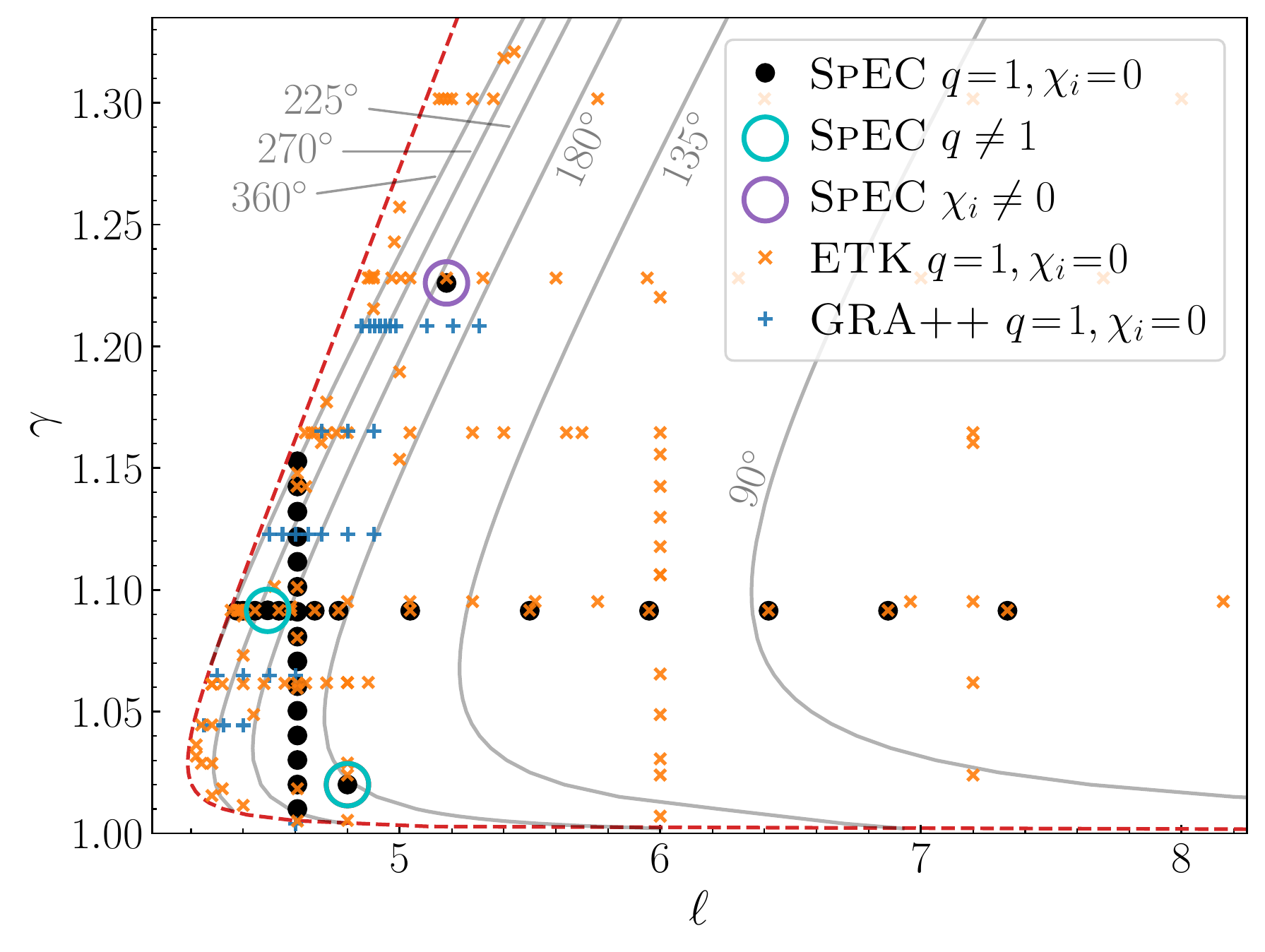}
\caption{ 
  Parameter space coverage of simulations:
  `\spec $q\!=\!1,\chi_i\!=\!0$' indicates equal-mass, non-spinning simulations presented
  here, and the large open circles denote sequences of \spec{} simulations
  of unequal masses (up to $q\!=\!10$), or with non-zero spins (up to
  $\chi_i\!=\!0.60$).
  The grey solid lines show contours of constant scattering angle
  generated using \SEOBfourPM{}, delineated by a fit
  of the scatter-capture separatrix (red dashed line)~\cite{Kankani:2024may}.
  For context, the plot also shows parameters of equal-mass, non-spinning BBH scattering simulations performed with \ETK{}~\cite{Damour:2014afa,Hopper:2022rwo,
  Rettegno:2023ghr,Swain:2024ngs} and \textsc{GRA++}~\cite{Albanesi:2024xus}.
  Ref.~\cite{Swain:2024ngs} also presents simulations
  extending up to $\gamma=1.96$, which are not shown on this plot.
}\label{Fig:ParameterSpace}
\end{figure}

  There are three NR codebases that have produced unbound BBH simulations and reported scattering angles:
  the Einstein Toolkit (\ETK{})~\cite{Damour:2014afa,Hopper:2022rwo,Rettegno:2023ghr,Swain:2024ngs},
  \textsc{GR-Athena++} (\textsc{GRA++})~\cite{Albanesi:2024xus}, and now \spec{}.
  The majority of these simulations have focussed on the equal-mass, non-spinning case.
  In this work, we present 60 \spec{} simulations across six slices of parameter space: two sequences for equal-mass non-spinning BHs,
  two sequences for equal-mass BHs with non-precessing spins (i.e., BH spin vectors parallel or antiparallel to the orbital angular momentum), and two sequences for unequal-mass non-spinning BHs.  Figure~\ref{Fig:ParameterSpace} shows the parameter-space coverage of the simulations presented here, as well as, those of the equal-mass non-spinning results available in the literature.
  One of our equal-mass non-spinning sequences reproduces a set produced by \ETK{} to
  allow for the cross validation between codes presented in Sec.\ \ref{sec:CompareToOtherNR}.
 Our simulations used a CPU time of $\sim 3.5$ million core-hours, which includes three resolutions for each simulation.

\subsection{Extracting the scattering angle}
\label{sec:ExtactScatteringAngle}

  The BH trajectories of all simulations discussed here are planar,
  and by virtue of initial conditions are restricted to the $xy$-plane.  Therefore, the aymptotic initial and final direction of motion of a
  BH is specified by azimuthal angles $\varphi^{\infty,{\rm in}}$ and $\varphi^{\infty,{\rm out}}$, respectively.
  The scattering angle $\theta_i$ of the $i$th BH 
is defined in terms of the asymptotic values of the azimuthal 
angles by
\begin{equation}
\label{eq:def_scattering_angle}
\theta_i = \varphi^{\infty,{\rm out}}_i - \varphi^{\infty,{\rm in}}_i - \pi.
\end{equation}
In order to account for the wraparound of trajectories, as in the lower panels of Fig.~\ref{Fig:Orbits}, we use the unwrapped azimuthal angles to define the asymptotic values.

The motion of each BH is tracked as the Cartesian coordinates of the center of the excised regions in the initial center-of-mass frame of the system. Each trajectory is split into incoming and outgoing legs and converted into polar $(r,\varphi)$ coordinates in the $xy$-plane.

The finite size of the numerical domain means that in order to get an accurate value of the scattering angle we need to extrapolate the trajectories to
infinite separation of the BHs. One way to do this is to fit $\varphi$ to a polynomial of the form
\begin{equation} \label{eq:PolyFit}
\varphi  = \varphi^\infty + \frac{\varphi^{(1)}}{D} + \frac{\varphi^{(2)}}{D^2} + \frac{\varphi^{(3)}}{D^3},
\end{equation}
where $D$ is the separation of the BHs. Previous works \cite{Damour:2014afa,Damour:2022ybd,Hopper:2022rwo,Rettegno:2023ghr,Albanesi:2024xus,Swain:2024ngs,Fontbute:2025vdv} varied the order of the polynomial whereas here we use a cubic polynomial in $1/D$, which we found to provide a good fit to the data without overfitting.

Alternatively, we can utilize the Keplerian relation between the separation and azimuthal angle
\begin{equation}\label{eq:Kepler}
D = \frac{p}{1+e\cos(\varphi-\varphi_0)},
\end{equation}
where $e$ is the eccentricity, $p$ is the semi-latus rectum, and $\varphi_0$ is a constant offset. To allow for easier fitting to the data, we rewrite \Eq{Kepler} using the trigonometric angle addition formula as
\begin{equation} \label{eq:KeplerFit}
\frac{1}{D} = A \cos \varphi + B \sin\varphi + C,
\end{equation}
and then fit the constants $A$, $B$, and $C$. We can return to the variables of Eq.~(\ref{eq:Kepler}) by
\begin{subequations}
\begin{align}
p =&\: \frac{1}{C},\\
e =&\: \frac{\sqrt{A^2+B^2}}{C},\\
\varphi_0 =&\: \text{arctan} (A/B),
\end{align}
\end{subequations}
and then determine the asymptotic angle via
\begin{equation}\label{eq:KeplerAngle}
\varphi^\infty = \varphi_0 \pm \arccos(-1/e),
\end{equation}
where we use $+$ for the ingoing leg and $-$ for the outgoing leg.

In order to obtain a value of the scattering angle with comprehensive 
errors, we fit the data to both the polynomial and Keplerian models, 
Eqs.~\eqref{eq:PolyFit} and~\eqref{eq:KeplerFit}, respectively, across
 a variety of fitting ranges in $D$.  The upper end of the fitting range is chosen 
   such that there is enough data to provide a good fit while the lower
  bound is limited to avoid using data in the strong field where the 
  functional forms of the fitting models break down. For the ingoing 
  legs, we restrict the fitting range to $D\in[d_{\rm in}, 200M]$ 
  with $d_{\rm in}\in[50M, 80M]$. Restricting the upper bound of the 
  fitting avoids including the junk radiation in the fitting data. 
  For the outgoing legs, we use $D\in[d_{\rm out}, 350M]$ with 
  $d_{\rm out}\in[80M, 230M]$. 

\begin{figure}[tb]
\centering
\includegraphics[width=\linewidth]{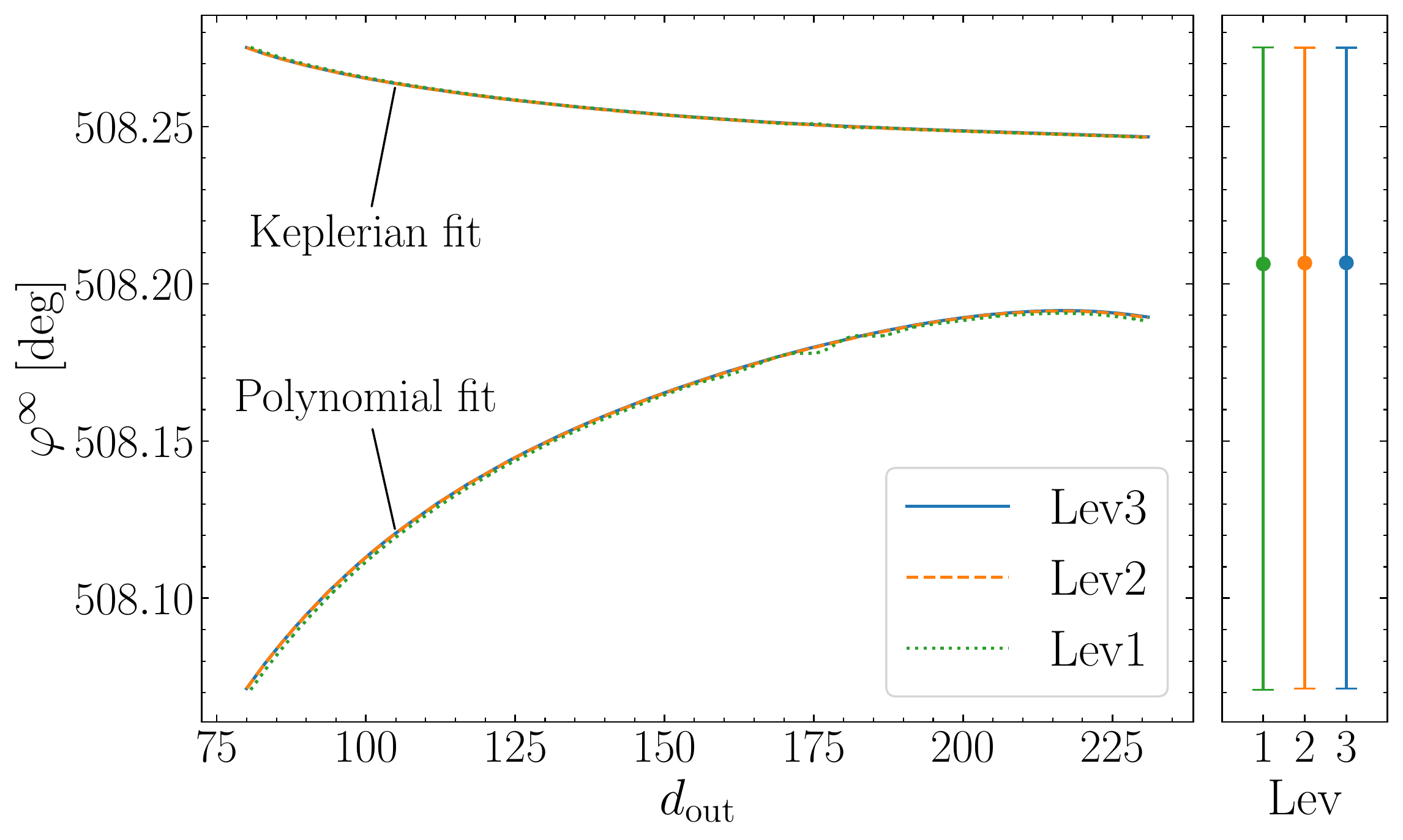}
\caption{Calculation of the asymptotic scattering angle.  The
    left panel shows the extrapolated outgoing angle obtained by fits
    over varying intervals $[d_{\rm out}, 350M]$.  The Keplerian fit
    Eq.~(\ref{eq:KeplerFit}) yields more stable results than the
    polynomial fit Eq.~(\ref{eq:PolyFit}).  The right panel indicates
    the asymptotic scattering angle calculated from both fits in the
    left panel, and our choice of error bar.  Both panels show three
    different resolutions (Lev1 through Lev3), which all give
    consistent results.}
\label{Fig:Fitting}
\end{figure}

For each range, we extrapolate the angle to $D\rightarrow \infty$ using
Eqs.~\eqref{eq:PolyFit} and~\eqref{eq:KeplerAngle}.  The value and error 
of the angle is taken as the average and range of the two models across 
the range of fitting data respectively. Figure~\ref{Fig:Fitting} demonstrates 
the results of the fitting across the different ranges of data and how 
we determine the errors. It shows the value of the extrapolated angle as 
a function of fitting range for both models across multiple resolutions 
(main plot) as well as the values of the extrapolated angle and its 
errors (right plot).

The dominant error in all of the scattering angle is the variance in 
the extrapolation values due to the range and model used.
For large fitting ranges the two methods deviate more, because the
fitting range includes more of the strong field region where the
approximations get worse.
The relative error resulting from the 
fitting is $\sim 10^{-4}$. 
Another source of error is the numerical error in the simulation. 
This is estimated by comparing the results of the different 
resolutions (Levs) and is found to be negligible (see the right 
plot in \Fig{Fitting}).

\subsection{Comparison between NR codes}
\label{sec:CompareToOtherNR}

In order to validate the results of our scattering simulations, we compare 
the scattering-angle values obtained from \spec{} with those calculated using 
\ETK{}~\cite{Swain:2024ngs}. 
The extrapolation of the scattering angles from the \ETK{} data is detailed 
in Sec.\ II.\ of Ref.~\cite{Swain:2024ngs}. This work uses the
  polynomial fitting procedure, Eq.~(\ref{eq:PolyFit}), with the fitting range fixed to $D\in[20M, 90M]$ 
for the ingoing leg and $D\in[30M, 180M]$ for the outgoing leg. Additionally, the 
polynomial order of the fitting is varied. The value of the angle corresponds to 
the highest polynomial order used and the error is then determined by taking the 
difference between this value and those obtained from fits using lower 
polynomial orders.

\begin{figure}[htb]
  \centering
  \includegraphics[width=\linewidth]{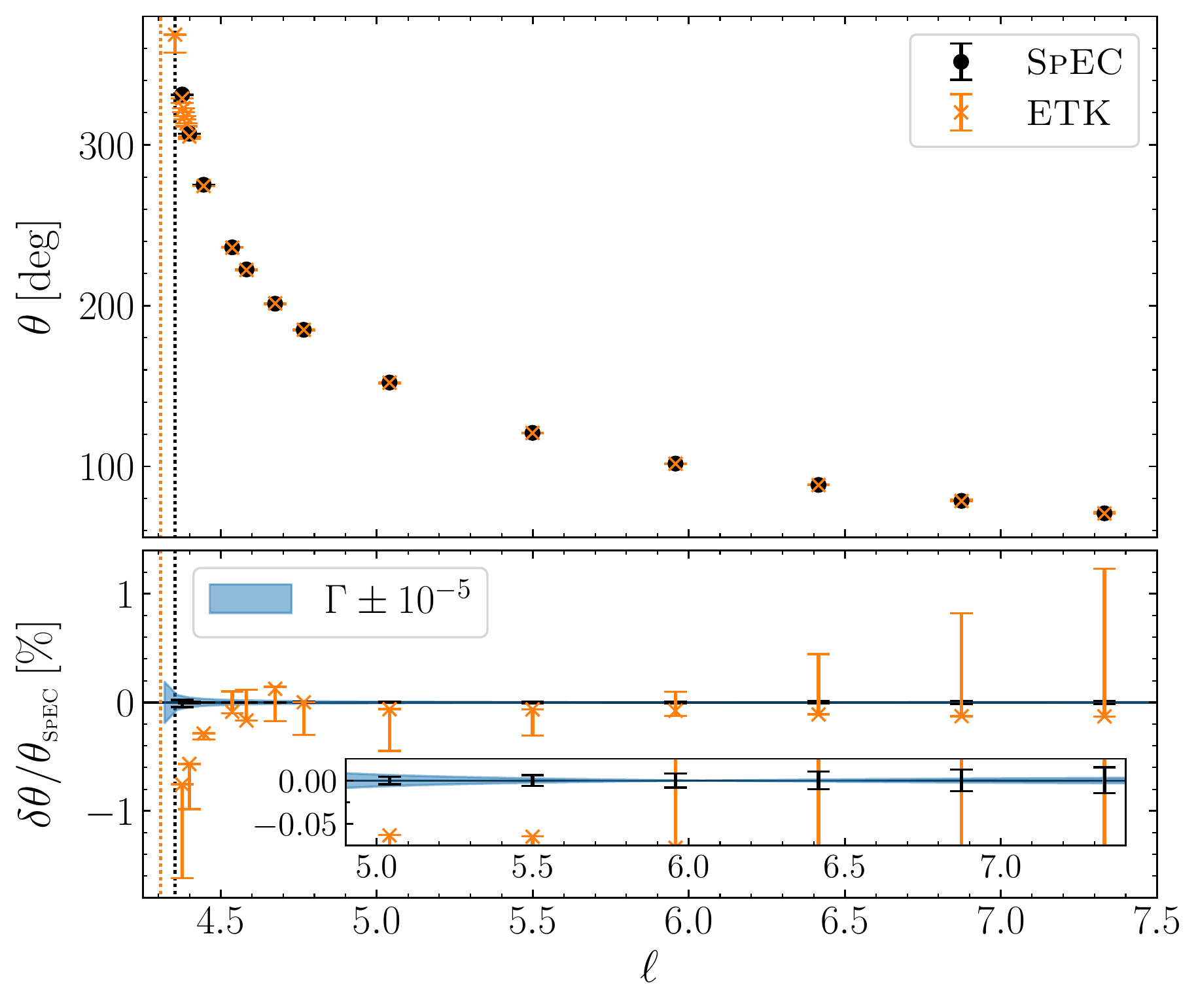}
  \caption{Comparison of scattering angle extracted from \spec{} and \ETK{}~\cite{Swain:2024ngs} simulations. 
  All simulations start as equal mass, non-spinning binaries with $\Gamma = 1.02264$.
  The top panel shows the relative scattering angle as a function of the initial angular momentum.
  The bottom panel shows the relative difference between the two codes. The \spec{} values lie
  on $\delta\theta=0$ with the error bars being the errors of the upper panel rescaled
  by the value of the angle. These errors are too small to be resolved on this scale, 
  but are resolved in the inset, which shows a zoom-in of the six largest 
  $\ell$ values.
  The blue shaded
  region on the lower plot demonstrates the expected difference in the angle due to a 
  $\pm 10^{-5}$ difference in the initial energy.
  The vertical dotted lines show the first confirmed capture from each code.
  }
    \label{Fig:ETKComparison}
  \end{figure}

We simulated a subset of the equal-mass, non-spinning dataset with $\Gamma \!=\! 1.02264$ $(\gamma\!=\!1.0916)$
from Ref.~\cite{Swain:2024ngs} and compare the results in \Fig{ETKComparison}. 
Because of symmetry under exchange of the two objects, the scattering angles for either BH are equal.  
To reduce numerical errors, the upper plot shows the average scattering angle, defined as
  \begin{equation}\label{eq:RelAngle}
    \theta_{\rm avg} = \frac{\theta_1 + \theta_2}{2},
  \end{equation}
where $\theta_{1/2}$ are the scattering angles of each BH, as defined in \Eq{def_scattering_angle}.
 Both codes exhibit the same divergent
behavior as the BHs approach the scatter-capture separatrix.
We note that the leftmost point ($\ell = 4.3536$) only has a data point for \ETK{} 
as the \spec{} simulation produced a dynamical capture. In Table II of 
Ref.~\cite{Swain:2024ngs} this run is marked as a potential dynamical capture due
 to the uncertainty in the scatter-capture separatrix and the fact that the BHs may
  merge if allowed to evolve to larger radii.  Indeed, after the initial encounter, the \spec simulation results in the BHs on a bound orbit with large apastron of $D=227.6 M$ before 
  merging during the secondary interaction. The \ETK{} simulations are terminated 
  at an outgoing separation of $D=180 M$ and hence the capture would not have been observed.

The lower panel of Fig.~\ref{Fig:ETKComparison} shows the relative difference between the two codes
\begin{equation}\label{eq:relativeDifference}
\frac{\delta\theta}{\theta_{\spec{}}} = \frac{\theta_{\ETK{}} - \theta_{\spec{}}}{\theta_{\spec{}}},
\end{equation}
where $\theta_{\ETK{}}$ and $\theta_{\spec{}}$ are the scattering angles from the respective codes.
The errors of the \spec{} results are significantly smaller than those from the \ETK{}
simulations, to the extent that they cannot be resolved on the same scale.
This is primarily due to the larger range of separation values used in the fitting of the \spec{} data.

The values from both codes show good agreement with each other, differing by less than $1\%$ 
relative difference. However, the scattering angles reported by either code do not always lie within 
each other's error bars. Both codes report the initial energy and angular momentum with an 
accuracy of approximately one part in $10^5$ (see Appendix \ref{App:Junk} for a discussion of the \spec{} accuracy).
 To demonstrate that this level of uncertainty is not responsible for the observed difference, 
 we show the expected variation in the scattering angle resulting from a $10^{-5}$ change in 
 the initial energy as the blue shaded region in the lower plot. This is obtained by varying 
 the initial energy of the \SEOBfourPM{} model (to be introduced in Sec.~\ref{sec:ClosedFormEOB}) by 
 $\pm 10^{-5}$ and computing the corresponding change in the scattering angle. The observed 
 difference $\theta_{\ETK}-\theta_{\spec}$ at small $\ell$ is significantly larger than what can be attributed 
 to variations in the initial conditions.

\section{Effective-One-Body models for scattering}
\label{sec:EOB}

In this section we introduce the families of EOB models that will be compared with 
NR scattering data. As discussed in the introduction, we split these into two categories: ``evolution" models, which are
based on evolving the EOB Hamiltonian and include the effects of radiation reaction through the
equations of motion, and ``closed-form'' models, which construct a prediction of the 
scattering angle via the so-called impetus formula.

\subsection{Evolution EOB models}

In this section we outline the building blocks used in the construction of the latest 
generation of \texttt{SEOBNR} models~\cite{Khalil:2023kep,Pompili:2023tna,
Ramos-Buades:2023ehm,vandeMeent:2023ols,Mihaylov:2023bkc,Gamboa:2024hli,Buonanno:2024byg}.
 In particular we describe the dynamics of the spinning non-precessing model based on 
the PN \texttt{SEOBNRv5HM}~\cite{Pompili:2023tna} Hamiltonian. We continue by 
 detailing the modifications necessary to adapt the \texttt{pySEOBNR} framework to be able 
 to accurately describe scattering binary configurations.

\subsubsection{EOB dynamics}
The EOB formalism~\cite{Buonanno:1998gg,Buonanno:2000ef,Damour:2000we,Damour:2001tu,Buonanno:2005xu} maps the two-body dynamics on to the effective dynamics of a test body in a deformed Schwarzschild or Kerr background with the deformation parametrized by the symmetric mass ratio $\nu$. The description of the conservative dynamics relies on a Hamiltonian $H_{\rm EOB}$, constructed through an effective Hamiltonian $H_{\rm eff}$ of a test mass $\mu$ moving in a deformed Kerr spacetime of mass $M$, and the following energy map connecting $H_{\rm eff}$ and $H_{\rm EOB}$:
\begin{equation}
H_{\rm EOB} = M \sqrt{1+2 \nu \left(\frac{H_{\rm eff}}{\mu}-1 \right)}\,.
\label{eq:eq01}
\end{equation}
The deformation of the Kerr Hamiltonian is obtained by imposing that at each PN order, the expanded EOB Hamiltonian agrees with a PN Hamiltonian through a canonical transformation. The dynamical variables of the generic EOB Hamiltonian are the orbital separation $\bm r$, the corresponding canonically conjugate momentum $\bm p$, and the spins $\bm{S}_{1,2}$.

The dissipative effects enter the dynamics through the radiation-reaction force $\mathcal{F}$, which are computed as a summation of PN GW modes (augmented with gravitational self-force information~\cite{vandeMeent:2023ols}) in a factorized form~\cite{Damour:2008gu, Damour:2007xr, Pan:2010hz, Damour:2007yf,Pompili:2023tna}.  For arbitrary orientations of the spins, both the spins and the orbital plane precess around the total angular momentum of the system $\bm J$. However, in this work we restrict to spins aligned with the orbital angular momentum of the systems. For such non-precessing spins the equations of motion read~\cite{Buonanno:2005xu,Khalil:2023kep}
\begin{equation}
\label{eq:fullSpin}
\begin{aligned}
\dot{r}&=\xi(r) \frac{\partial H_{\rm EOB}}{\partial p_{r_*}}, \quad
&\dot{\phi} &=\frac{\partial H_{\rm EOB}}{\partial p_{\phi}},\\
\dot{p}_{r_*}&=-\xi(r)\frac{\partial H_{\rm EOB}}{\partial r} +\mathcal{F}_{r}, \quad
&\dot{p}_{\phi}&=\mathcal{F}_{\phi},
\end{aligned}
\end{equation}
where the evolution of the radial momentum employs the transformation to tortoise-coordinates, $p_{r_*} = p_r \xi(r)$,  with $\xi(r) = dr/dr^*$ (see Refs.~\cite{Taracchini:2013rva,Pan:2013rra} for details).

\subsubsection{\SEOBNR{} Hamiltonian}

For the \texttt{SEOBNRv5} family~\cite{Khalil:2023kep,Pompili:2023tna,Ramos-Buades:2023ehm,vandeMeent:2023ols,Mihaylov:2023bkc,Gamboa:2024hli}
 of inspiral-merger-ringdown \texttt{SEOBNR} models, the Hamiltonian is given in Ref.~\cite{Khalil:2023kep}, and it reduces as $\nu\to 0$ to the Kerr Hamiltonian for a test mass in a generic orbit. In the non-precessing case we use the Hamiltonian of the accurate multipolar aligned-spin \texttt{SEOBNRv5HM} model~\cite{Pompili:2023tna}, which includes 4PN information for spinning binaries and arbitrary mass ratios, 
\begin{subequations}
\begin{align}
    \label{eq:HeffAnzAlign}
    H_\text{eff} &=H_{\rm even} + H_{\rm odd}, \\
    H_{\rm even} &=  \Bigg[ A^\text{align} \bigg( 
    \mu^2 + p^2 + B_{np}^\text{align} p_r^2  \nonumber \\ 
    & \qquad + B_{npa}^\text{Kerr\,eq}  \frac{p_\phi^2 a_+^2}{r^2} + Q^\text{align}
    \bigg)\Bigg]^{1/2}\,,\\
    H_{\rm odd} &= \frac{Mp_\phi \left(g_{a_+} a_+ + g_{a_-} \delta a_-\right) + \text{SO}_\text{calib}
    + G_{a^3}^\text{align}}{r^3+a_+^2 (r+2M)},
\end{align}
\end{subequations}
where $a_\pm = a_1 \pm a_2$. The two terms in the Hamiltonian include the even-in-spin and the odd-in-spin contributions, respectively. 
The term  $H_{\rm even}$ contains the non-spinning and spin-spin contributions through the potentials $A^\text{align}(r)$, $B_{np}^\text{align}(r)$ and $Q^\text{align}(p_{r_*},r)$, with no $S^4$ corrections needed since the Kerr Hamiltonian reproduces all even-in-spin leading PN orders for BBHs~\cite{Vines:2016qwa}. The potentials include 4PN spin-spin information,
\begin{subequations}
\begin{align}
A^\text{align} &= \frac{a_+^2/r^2+A_\text{noS}+A_\text{SS}^\text{align}}{1+(1+2M/r)a_+^2/r^2}, \\
B_{np}^\text{align} &= -1 + \frac{a_+^2}{r^2} + A_\text{noS} \bar{D}_\text{noS} + B_{np,\text{SS}}^\text{align}, \\
B_{npa}^\text{Kerr\,eq} &= -\frac{1+2M/r}{r^2+a_+^2 (1+2M/r)},\\
Q^\text{align} &= Q_\text{noS} + Q_\text{SS}^\text{align},
\end{align}
\end{subequations}
where the $A$, $B$, and $D$ potentials are functions of $r$ and the $Q$ potentials depend on $r$ and $p_{r_*}$.
The $A_\text{noS}$ term is Taylor-expanded up to 5PN in $u\equiv M/r$ with the replacement of the coefficient $u^6$ (except for the log term), by the parameter $a_6$ which is calibrated to NR simulations. The explicit expressions for the potentials can be found in Ref.~\cite{Khalil:2023kep} and in Appendix A of Ref.~\cite{Pompili:2023tna}.

The odd-in-spin term, $H_{\rm odd}$, contains the gyromagnetic factors, $g_{a_+}(p_\phi,r)$ and $g_{a_-}(p_\phi,r)$, expanded up to 3.5PN~\cite{Khalil:2023kep}, the function $G_{a^3}^\text{align}(p_\phi,r)$ which includes $S^3$ corrections (see Appendix A of
Ref.~\cite{Pompili:2023tna} for the explicit expressions) and a calibration term at 4.5PN, 
\begin{equation}
	\label{eq:SO_calib}
	\text{SO}_\text{calib} =\nu d_\text{SO} \frac{M^4}{r^3} p_\phi a_+.
\end{equation}
to increase the accuracy against NR waveforms.

\subsubsection{Modifications for scattering}
\label{sec:EOBMods}

The \SEOBNR{} model is designed to model the inspiral, merger, 
and ringdown of BBHs. In order for the models to be 
applicable to the scattering case, we need to modify the implementation 
within the \texttt{pySEOBNR} framework. The primary modification involves
the initial conditions where we supply the initial 
energy, angular momentum, mass ratio, (aligned) spins, and initial separation. 
These need to be converted to the parameters the model evolves. The only 
parameter we do not have immediate access to is the radial momentum.
We obtain it through a numerical one-dimensional root-finding such 
that the Hamiltonian equals the initial energy.

We also modify the termination conditions to stop the evolution 
when the BHs reach a separation of $D_0+100M$.
The scattering angle is 
calculated as the difference between the azimuthal angles of the BHs at the start and at
the end of the simulation. We choose an initial separation $D_0=10^6 M$, which 
yields a relative error in the scattering angle due to the $D>D_0$ 
contribution of less than $10^{-6}$.

  One advantage of the evolved models is that they can also simulate dynamical capture scenarios. This allows them to determine the critical parameters where the system is captured as well as the maximum value of the scattering angle on the separatrix. 
This is in contrast to the closed-form models where there is no maximum 
value of the scattering angle,
only the divergence of the model close to the separatrix.  

\subsection{Closed-form EOB}
\label{sec:ClosedFormEOB}

In contrast to the evolved models, closed-form resummation models have been introduced specifically for predictions of the scattering angle.
These include the $w_{\rm EOB}$ model~\cite{Rettegno:2023ghr,Damour:2022ybd} (see also Refs.~\cite{Kalin:2019rwq,Kalin:2019inp}) and the \SEOBPM{} model~\cite{Buonanno:2024vkx} to which we will compare NR results in this work.
The closed-form models describe the two-body dynamics using an effective potential $w$,
\begin{align}\label{eq:impetus}
    p_r^2
    &=
    \pin^2
    -
    \frac{L^2}{r^2}
    +w(E_{\rm eff},L,r;a_\pm)
    \,,
\end{align}
where $w\to0$ as $r\to\infty$, and so $p_r\to\pin$.
The \textit{impetus formula}~\eqref{eq:impetus} (see e.g.~Refs.~\cite{Kalin:2019inp,Kalin:2019rwq}) is an inverse Hamiltonian,
wherein the radial momentum $p_r$ is expressed as a function of $E_{\rm eff}$ and the other kinematic variables,
instead of $E_{\rm eff}=H_{\rm eff}(p_r,L,r;a_\pm)$.
Such a starting point is advantageous as the scattering angle, which we seek to describe,
is given straightforwardly by the Hamilton-Jacobi formalism:
\begin{align}\label{eq:angleFromPR}
    \theta+\pi
        =
        -2\int_{r_{\rm min}}^\infty\!{\rm d}r\frac{\partial}{\partial L}
        p_{r}
        \,,
\end{align}
$r_{\rm min}$ being the point of closest approach between the two scattered bodies, the solution to $p_r(r_{\rm min})=0$.

  The relationship~\eqref{eq:angleFromPR} demonstrates a one-to-one correspondence between weak-field perturbative components of the scattering angle $\theta$ and the potential $w$.
  Thus, by matching perturbative PM results for the two-body scattering angle with the potential through Eq.~\eqref{eq:angleFromPR}, one guarantees consistency of these models with the perturbative regime.
  This matching of the gauge-invariant scattering angle avoids the need for canonical transformations and only in the construction of a suitable ansatz for the potential does its gauge dependency show up.

The main difference between the two closed-form models \SEOBPM{} and $w_{\rm EOB}$ is the choice of ansatz for $w$.
First, the $w_{\rm EOB}$ model uses a simple perturbative ansatz,
  \begin{align}\label{eq:wEOBPerturbative}
    w_{n\rm PM}
    &=
    \sum_{m=1}^{n} u^m
    w^{(m)}\,,
  \end{align}
with perturbative PM coefficients $w^{(m)}$ that are functions only of the effective energy $E_{\rm eff}$, masses and spins.
This model is not an EOB model in the sense that its $\nu\to0$ limit does not reproduce scattering in a Kerr background,
thus we will henceforth refer to this model as $w_{n{\rm PM}}$.
However,
while this potential is, from that perspective, perturbative, 
the angle calculated via Eq.~\eqref{eq:angleFromPR} is nevertheless a
resummation of the perturbative PM angle.

The \SEOBPM{} model takes as its starting point the motion of a probe $\mu$ moving
in an effective background metric $g_{\rm eff}^{\mu\nu}p_\mu p_\nu=-\mu^2$.
By doing so, it captures not only the perturbative weak-field limit
but also the extreme-mass ratio $\nu\to0$.
Solving this on-shell condition for the radial momentum,
\begin{subequations}
    \begin{align}
      p_r^2
      &=
      \frac1{(1+B_{\rm np}^{\rm Kerr})}\bigg[\frac1{A}\left(E_{\rm eff}-\frac{ML(g_{a_+}a_++g_{a_-}\delta \: a_-)}{r^3+a_+^2(r+2M)}\right)^2
      \nonumber\\
      &\qquad-\left(\mu^2+\frac{L^2}{r^2}+B_{\rm npa}^{\rm Kerr}\frac{L^2a_+^2}{r^2}\right)\bigg]\,,\\
      & \equiv
      \pin^2-\frac{L^2}{r^2}
      +
      w_{\rm \SEOBPM{}}(E_{\rm eff},L,r;a_\pm),
    \end{align}
\end{subequations}
this implicitly defines the effective potential $w_{\rm \SEOBPM{}}$.
The function $A$ and the gyro-gravitomagnetic factors $g_{a_\pm}$ are
\begin{align} \label{eq:def}
  A&=A^{\rm Kerr}+\frac{\Delta A}{1+\chi_+^2u^2(2u+1)}\,,&
  g_{a_\pm}&=\frac{\Delta g_{a_\pm}}{u^2}
  \ ,
\end{align}
and $A^{\rm Kerr}(r)$, $B_{\rm npa}^{\rm Kerr}(r)$, and $B_{\rm np}^{\rm Kerr}(r)$ are all parameters of the Kerr metric.
Perturbative PM deformations are incorporated into $A(E_{\rm eff},\chi_\pm,r)$ and $g_{a_\pm}(E_{\rm eff},\chi_\pm,r)$ through $\Delta A(E_{\rm eff},\chi_\pm,r)$ and $\Delta g_{a_\pm}(E_{\rm eff},\chi_\pm,r)$ in a similar fashion to Eq.~\eqref{eq:wEOBPerturbative}.
Full details are given in Ref.~\cite{Buonanno:2024vkx}.

In both closed-form models, the PM deformation coefficients, $w^{(m)}$ and $\Delta A$, $\Delta g_{a_\pm}$, are uniquely determined by the perturbative PM coefficients of the scattering angle.
More precisely, they are polynomial functions of each other; the analytic complexity of the closed-form potentials is thus equivalent to the PM scattering angle.
The $n$PM scattering-angle coefficients have a simple form.
First, they are $\lfloor\tfrac{n-1}{2}\rfloor$SF exact where $n$SF denotes an expansion in the symmetric mass ratio $\nu$ to the $n$th order (i.e. 1PM and 2PM are 0SF exact, 3PM and 4PM are 1SF exact and so on).
Second, non-trivial analytic dependence appears only as functions of the effective energy $E_{\rm eff}$.
At leading 1PM and 2PM orders all dependence is rational, and at higher $n$PM orders multiple polylogarithms of weight $(n-2)$~\cite{Driesse:2024feo} appear (at 5PM order, the spinless 2SF part is still unknown).
Additionally at the 4PM order, complete elliptic integrals appear~\cite{Bern:2021dqo}.

While the closed-form models seem conservative by construction, one may include dissipative effects by considering different choices of $\theta$ in Eq.~\eqref{eq:angleFromPR}.
For a conservative model, one would use the conservative analytic PM angle.
However, one may also match the potentials against dissipative angles.
With dissipation, the effects of recoil generally imply that the scattering angles of each BH differ, and one may then consider different choices for $\theta$.
These include the relative scattering angle $\theta_{\rm rel}$~\cite{Bini:2021gat,Driesse:2024feo}
and the individual scattering angles $\theta_i$ of each BH.

The relative scattering angle commonly used in PN and PM literature is defined as the angle between incoming and outgoing relative momenta, $\boldsymbol{p}_{\rm in}$ and $\boldsymbol{p}_{\rm out}$, in the initial and final CoM frames.
That is, importantly, the outgoing spatial momentum $\boldsymbol{p}_{\rm out}$ is defined in the final CoM frame --- which differs from the initial CoM frame
due to radiation of four momentum of the two black holes.
Its computation thus involves the recoil vector $\boldsymbol{P}_{\rm recoil}$ describing the back-reaction of the CoM frame ---
for more detail, see e.g.~Refs.~\cite{Dlapa:2022lmu,Driesse:2024feo}.
The relative angle is then simply the angle between the incoming and outgoing momenta:
\begin{align}
  \cos\theta_{\rm rel}=
  \frac{\boldsymbol{p}_{\rm in}
  \cdot \boldsymbol{p}_{\rm out}}{
    |\boldsymbol{p}_{\rm in}||\boldsymbol{p}_{\rm out}|
  }
  \ .
\end{align}
For a covariant version of this formula we refer to Eq. (26) of Ref.~\cite{Jakobsen:2023pvx}.
The relative scattering angle is symmetric in the two BHs,
but it is also different from the averaged angle~\eqref{eq:RelAngle}:
\begin{align}\label{eq:averel}
    \theta_{\rm avg}-\theta_{\rm rel}
    =
    \frac{E_1-E_2}{2M\mu\sqrt{\gamma^2-1}}
    (\hat{\boldsymbol{b}}-\theta_{\rm rel}\hat{\boldsymbol{p}})
    \cdot
    \boldsymbol{P}_{\rm recoil}
    +
    \mathcal{O}(G^6)
    \ ,
\end{align}
with the discrepancy starting at 4PM.
Here, $\hat{\boldsymbol{b}}$ and $\hat{\boldsymbol{p}}$ are unit vectors in the directions of the inital impact parameter and relative momentum and $E_i$ are the initial energies of the $i$th BH.

  As a third option for $\theta$, one could also use the average angle.
  Physically, however, one might argue that it is most sensible to use an angle derived from the trajectory of the genuine BHs ($\theta_i$) or the approximate relative coordinates ($\theta_{\rm rel}$) as opposed to an average of two such angles.
  Further, we note that knowledge of the relative angle together with recoil and loss of energy uniquely fixes the individual angles.

Thus, for each potential, we have four subclasses of models depending on which angle they were matched against: a conservative potential matched from the conservative angle, a dissipative potential matched from the (dissipative) relative scattering angle and two asymmetric models, one for each $\theta_i$, matched from the individual scattering angles.
Further, the models are labeled by the precision to which they reproduce $\theta$.
Namely, for each model (conservative, relative and asymmetrical) we have a $n$PM version with $n\le5$.
Note, however, that dissipative effects start only at 3PM such that all angles agree until then.
Further, recoil effects start only at 4PM (unequal masses) and 5PM (unequal spins) such that the asymmetric models only distinguish themselves there.

The 5PM scattering angle is known only to the 1SF accuracy with its 2SF ($\nu^2$) piece still undetermined.
As such, the 5PM models considered in this work ---namely \SEOBfivePM{}--- are matched only at 1SF accuracy.
Technically, and for consistency, we include $G^5$ 2SF terms to 3rd PN order which are determined from the potential itself via the 4PM results (generally $n$PM fully determines $(n-1)$PN observables through the potential).
These 2SF 3PN terms must be included in the 5PM angle in order that the matched potential is free of poles in its PN limit.
More details on the \texttt{SEOB-5PM(1SF)} model are given in Appendix~\ref{App:5PMModel}.

The asymmetric models have not previously been considered and are introduced in this work to capture recoil effects on the scattering angle.
Again, there are two asymmetric models built to reproduce the scattering angle $\theta_i$ of each individual BH.
The potentials of each model are determined by inserting analytic $n$PM expressions for the individual scattering angle of the $i$th BH $\theta_i$ in Eq.~\eqref{eq:angleFromPR} and enforcing the validity of this equation to the $n$th PM order (for an $n$PM order model).
In particular, using these models we may compute a prediction for the angle difference $\theta_2-\theta_1$.
Namely, we compute $\theta_1$ using the first asymmetrical model and $\theta_2$ using the second one and calculate their difference.

Dissipative models based on the relative scattering angle accurate to the fourth (physical) PM order were constructed in Refs.~\cite{Rettegno:2023ghr,Buonanno:2024vkx} (with partial 5PM spin deformations).
A model including conservative 5PM 1SF deformations (based on the PM results of Ref.~\cite{Driesse:2024xad}) was considered in Ref.~\cite{Swain:2024ngs}.
However, the recent 5PM dissipative results of Ref.~\cite{Driesse:2024feo} have not yet been tested in EOB models.

\section{Scattering-angle comparisons}
\label{sec:CompareToEOB}

In this section we compare the scattering angles calculated from \spec{} with those obtained using the EOB models described in Sec.\ \ref{sec:EOB}, as well as the publically available evolution model 
 \Dali{}\footnote{In particular, we compare to \texttt{v.1.1.0} available at \url{https://bitbucket.org/teobresums/teobresums/src/v1.1.0-Dali/}.}. One major difference between the evolution models is that \SEOBNR{}
 only includes quasicircular information in the radiation-reaction terms,
 while \Dali{} includes analytical non-circular corrections to the 
 azimuthal radiation-reaction force~\cite{Nagar:2024oyk,Chiaramello:2020ehz}.
 Generally, unless otherwise stated, we use the closed-form models based on the relative angle which include all dissipative effects encoded in the relative angle.

\begin{figure*}
  \centering
  \includegraphics[width=0.98\linewidth,trim=0 8 0 0]{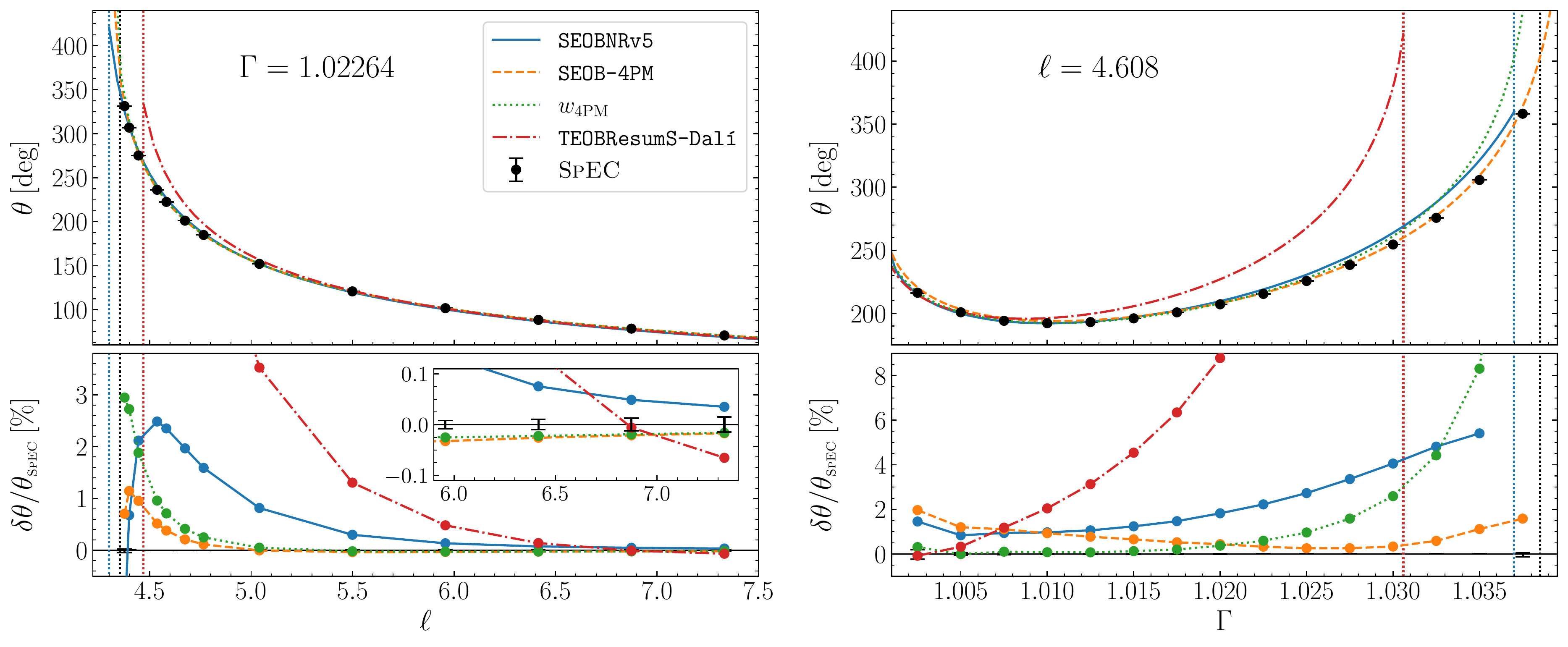}
  \caption{Comparison of EOB scattering angle results with those calculated by \spec{} for equal-mass non-spinning systems. 
  The top panels show the relative scattering angle for constant $\Gamma=1.02264$ ({\em left}) and constant $\ell=4.608$ ({\em right}).
  The bottom panels show the relative difference in the scattering angle as compared to the NR values with the inset in the left plot showing a zoom in of the four largest $\ell$ points. The vertical dotted lines show the first confirmed capture.
  }
  \label{Fig:EqualMass}
  \end{figure*}

For each simulation
we provide the initial conditions, \spec{} scattering angle (with errors), and the {\em relative}
scattering angle calculated using each EOB model in 
Appendix \ref{app:ScatteringAngles}. This data, as well as the 
exact initial parameters of the simulations, are also available in the ancillary data file.

In the comparison plots below,
we show the average value of the \spec{} scattering angles, $\theta_{\spec{}}$, 
and the relative scattering calculated using each EOB model, $\theta_{\rm EOB}$.
The evolution models, \SEOBNR{} and \Dali{}, do have errors arising from 
the finite initial and final separations; however, as discussed in 
Sec.~\ref{sec:EOBMods}, these are negligible on the scale of the plots.
We also present the relative difference to the \spec{} values, defined as
\begin{equation}
    \frac{\delta\theta}{\theta_{\spec{}}} = \frac{\theta_{\rm EOB} - \theta_{\spec{}}}{\theta_{\spec{}}}.
\end{equation}
For the unequal mass systems we also show the individual angles of both BHs and compare the difference in
the angles to that predicted by the \SEOBPM{} models.

\subsection{Equal-mass non-spinning black holes}

Here, we present two comparisons to sets of simulations,
one of constant $\Gamma =1.02264$ (the same as in Sec.~\ref{sec:CompareToOtherNR})
and the other of constant $\ell=4.8$. As the system is symmetric in the mass ratio and spins,
we only compare
the relative scattering angle as defined in \Eq{RelAngle}.  Our findings are summarized in \Fig{EqualMass}.

  The left panel of \Fig{EqualMass} shows the comparison of the 
  scattering angle for the sequence of simulations at constant energy $\Gamma$.
  Most of the EOB models show agreement with the NR results to 
  $<3\%$ right up to capture, which matches previous comparisons in the 
  literature~\cite{Damour:2014afa,Rettegno:2023ghr,Swain:2024ngs}.
  The exception to this is
  \Dali{} which diverges from the NR results as we decrease $\ell$ and predicts
  that the separatrix is at a larger $\ell$. In contrast, the \SEOBNR{} model stays 
  faithful to the NR results up to the NR capture point, but it predicts that the 
  systems stay unbound until a lower $\ell$ value. Due to the
  increased accuracy of the \spec{} results we can look at the
  behavior of the EOB models when approaching the weak field.
  The inset in the lower panel shows that the EOB models are all tending towards
  the NR values as we increase $\ell$ apart from \Dali{} which appears to slightly
  overshoot.

  The right panel of \Fig{EqualMass} shows the comparison of the 
  scattering angle for the sequence of simulations with constant angular momentum $\ell=4.8$.
  At low energies, all of the EOB models show good agreement with the NR results to 
  within $\sim 2\%$. However, as we increase the 
  energy, the accuracy of some of the models breaks down.
  In particular, \Dali{} quickly diverges from the NR results and predicts
  that the separatrix is at a significantly lower energy than the NR results.
  The other models show good agreement ($\lesssim 6\%$ difference) to the NR values
  apart from $w_{\rm 4PM}$, which diverges at the highest energies.

\begin{figure}
  \centering 
  \includegraphics[width=\linewidth,trim=0 8 0 0]{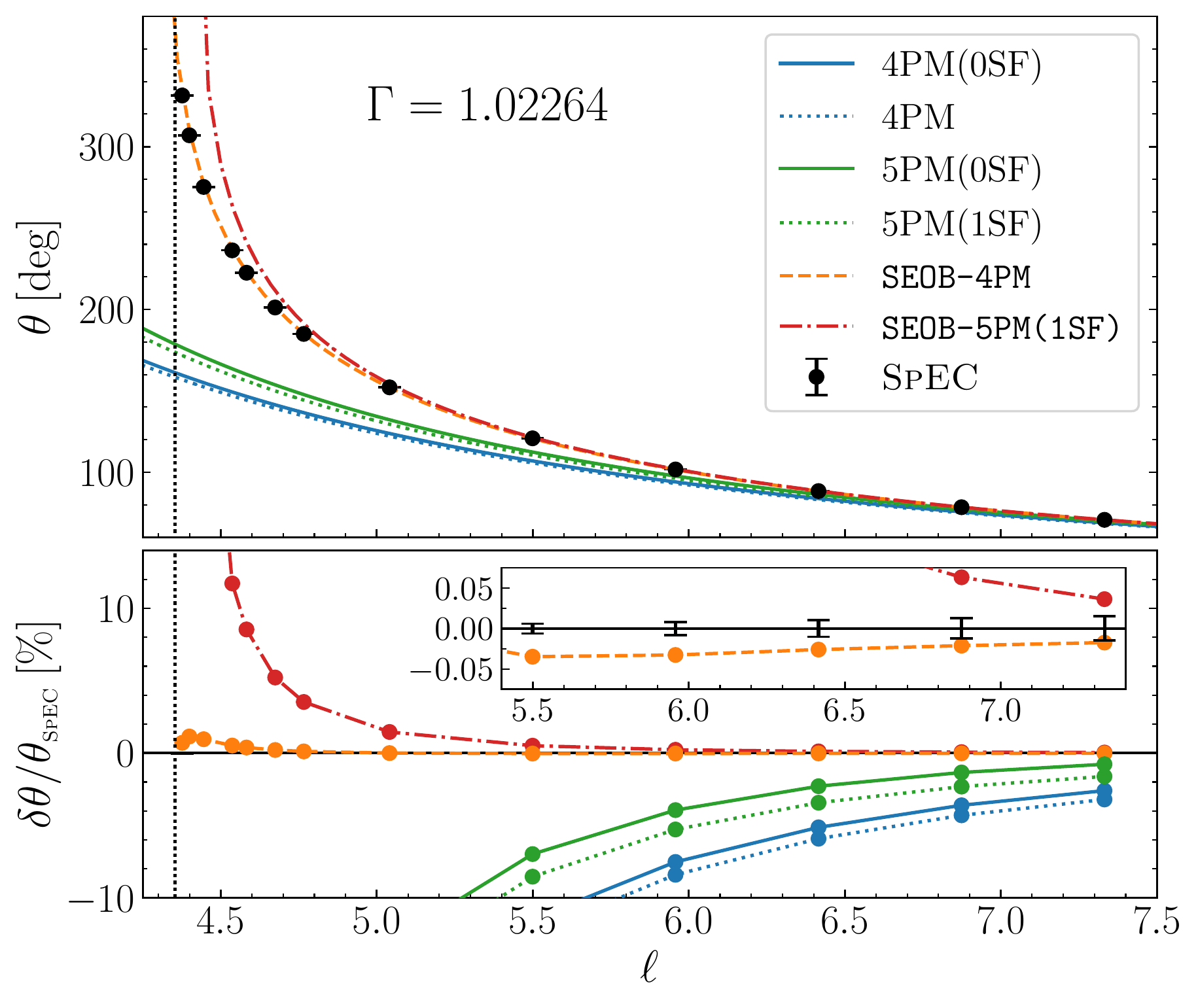}
  \caption{Scattering angle results for equal mass systems with $\Gamma = 1.02264$ and comparisons to higher order PM models. In the legend, $n$PM has all terms up to $G^n$ whereas $n$PM($m$SF) has all terms up to $G^{n-1}$ plus contributions to $G^n$ up to $\nu^m$. The vertical dotted line shows the first confirmed capture of the \spec{} simulations.
  }
    \label{Fig:HighOrderPMComparison} 
  \end{figure}

  In Fig.~\ref{Fig:HighOrderPMComparison}, we 
  explore the effects of higher order PM contributions by comparing
  the \SEOBfourPM{} and \SEOBfivePM{} models.
  Evidently, the agreement of the \SEOBfourPM{} with NR is best.
  This could be explained by the missing 5PM(2SF) contributions in the \SEOBfivePM{} model because, naively, all SF orders are similarly important for the equal-mass phase space considered here.
  In addition to the \SEOBPM{} models, we have also plotted the perturbative PM angles (i.e. without any resummation) at 4PM and 5PM precision including, either, only 0SF contributions (4PM(0SF) and 5PM(0SF)) or 0SF and 1SF contributions (4PM and 5PM(1SF)).
  Here, it seems that the 1SF contributions pull the (geodesic) 0SF angles away from the NR result worsening the agreement.
  However, when one considers exact 0SF results, the angle overshoots and from this perspective the 1SF contributions should decrease the angle slightly (see Fig.~4 of Ref.~\cite{Driesse:2024feo}).
  We also note the interesting feature, that at the perturbative level, 1SF corrections are very small compared with 0SF.
  However, the resummation models are very sensitive even to these very small variations.

\begin{figure*}
  \centering
  \includegraphics[width=0.98\linewidth,trim=0 5 0 0]{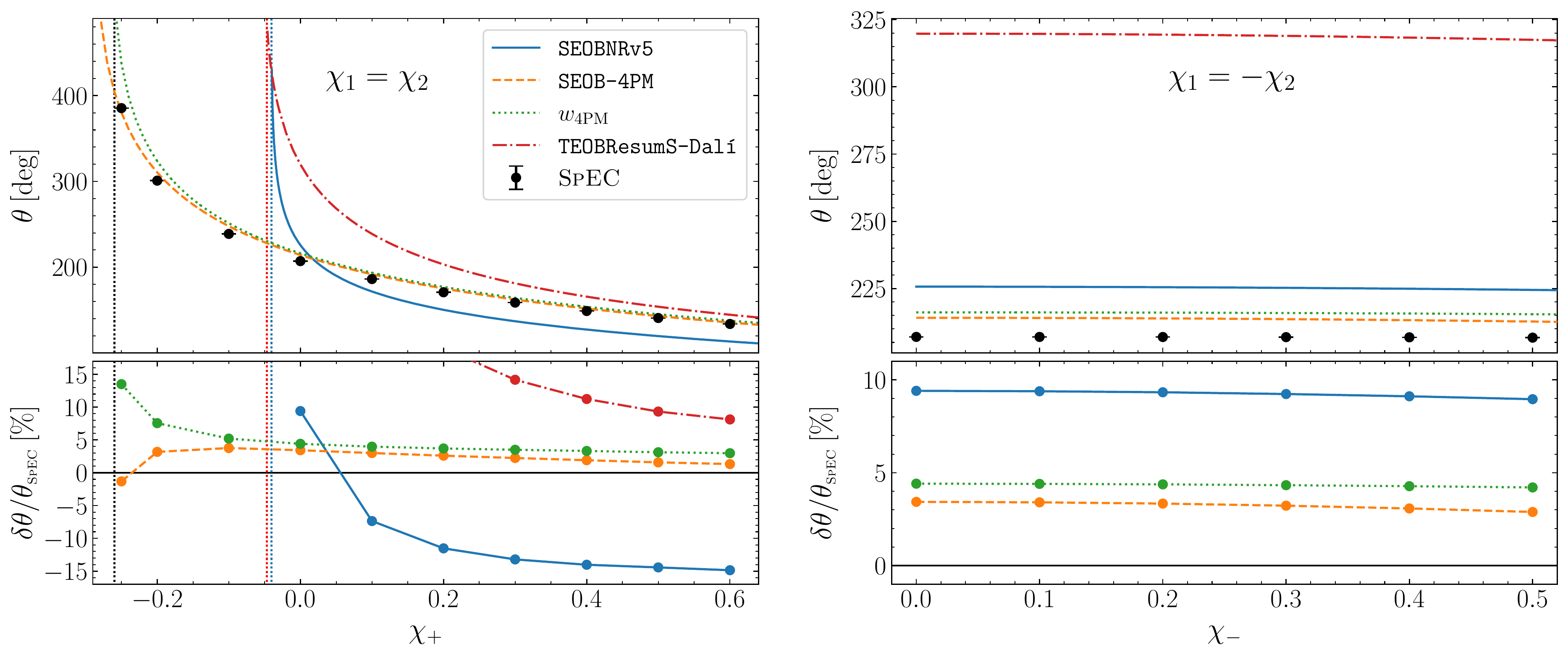}
  \caption{Comparison of EOB scattering angle results with those calculated by \spec{} for equal-mass aligned-spinning systems with constant $\Gamma = 1.055$ and $\ell = 5.18$. 
  The top panels show the relative scattering angle for equal parallel spins ({\em left}) and equal magnitude, anti-parallel spins ({\em right}) simulations.
  The bottom panels show the relative difference in the relative scattering angle as compared to the NR values. The vertical dotted lines on the left plot show the first confirmed capture.
  }
  \label{Fig:Spin}
  \end{figure*}  

\subsection{Equal-mass aligned-spin black holes} 

To study the effect of spins we
consider equal-mass systems with spins aligned with the orbital angular momentum. 
Our sets of simulations have constant $\Gamma = 1.055$ and $\ell = 5.18$ and include two spin configurations:
one with equal parallel spins ($\chi_1 = \chi_2$) and the other with equal-magnitude, anti-parallel spins ($\chi_1 = -\chi_2$).

\begin{figure}[b]
  \centering 
  \includegraphics[width=0.98\linewidth,trim=0 8 0 0]{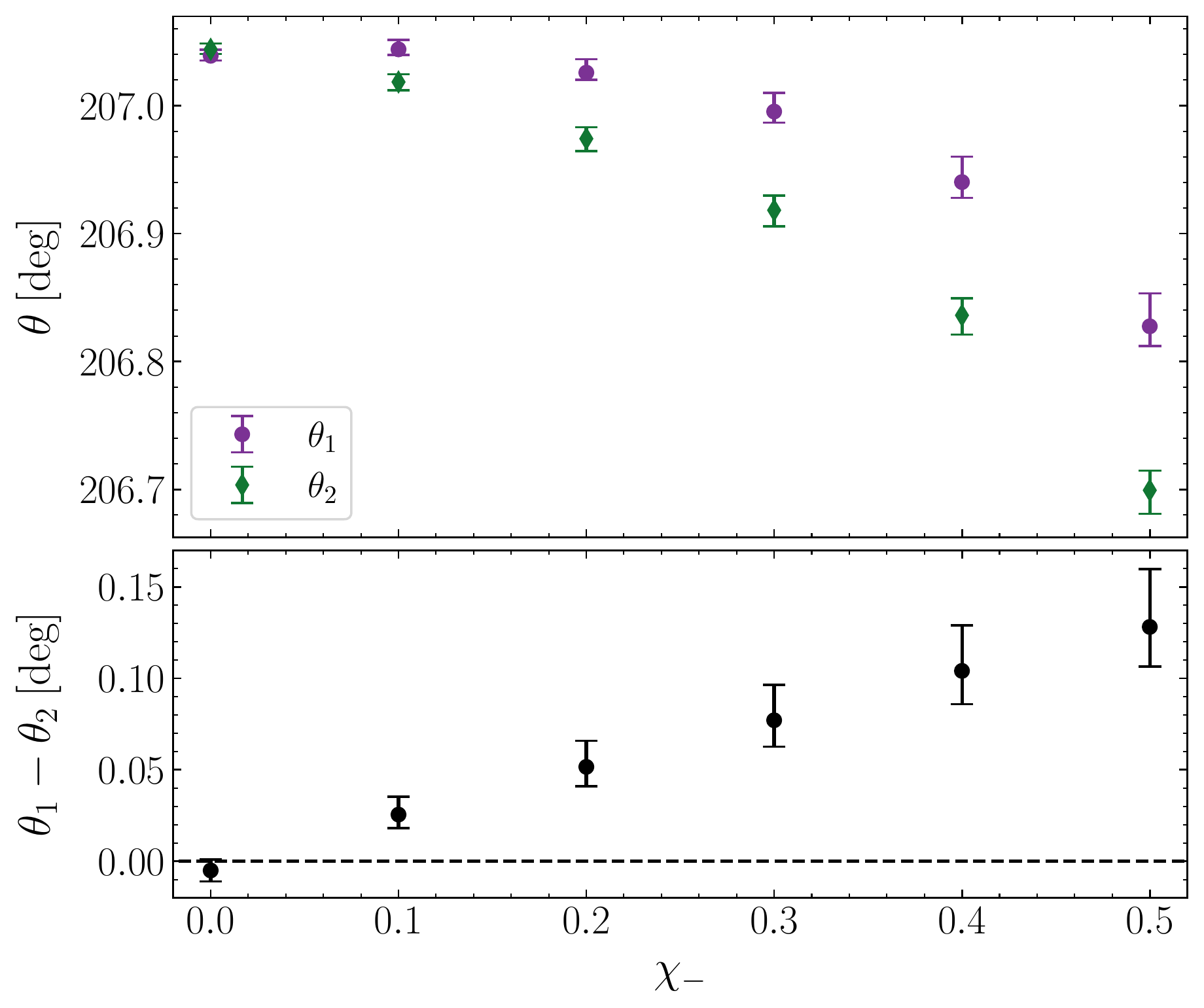}
  \caption{Scattering angle results for equal mass systems with $\Gamma = 1.055$ and $\ell = 5.18$,
  and equal magnitude anti-parallel spins ($\chi_+ = 0$).
  $\theta_2$ ($\theta_1$) represents the scattering angle of the BH with spin (anti-)parallel with the orbital angular momentum.
  The difference in the scattering angles is shown in the lower panel.
  }
    \label{Fig:AntiSpinDiff} 
  \end{figure}

  The left panel of \Fig{Spin} shows the 
  scattering angle as a function of spin for the equal-mass parallel-spin simulations for the four EOB models.
  The upper plot shows that decreasing the total angular momentum,
  by decreasing $\chi_+$, increases the scattering angle. 

  The closed-form models
  show a very similar behavior to the NR results and are generally within a few percent of the \spec{} values.
  The evolution models both predict a capture for $\chi_+ \lesssim -0.1$, which starts
  at significantly lower spin than the \spec{} capture range ${\chi_+ \lesssim -0.26}$.
  \SEOBNR{} agrees with the NR results in the non-spinning limit, whereas 
  \Dali{} tends towards the \spec{} values when increasing $\chi_+$.
  This behavior is consistent with the results of Ref.~\cite{Hopper:2022rwo} 
  where an earlier version of \texttt{TEOBResumS} also showed disagreement 
  with NR results exceeding 10\% for aligned-spin systems, as well as, predicting
  premature captures.

\begin{figure*}
  \centering
  \includegraphics[width=0.97\linewidth,trim=0 10 0 12]{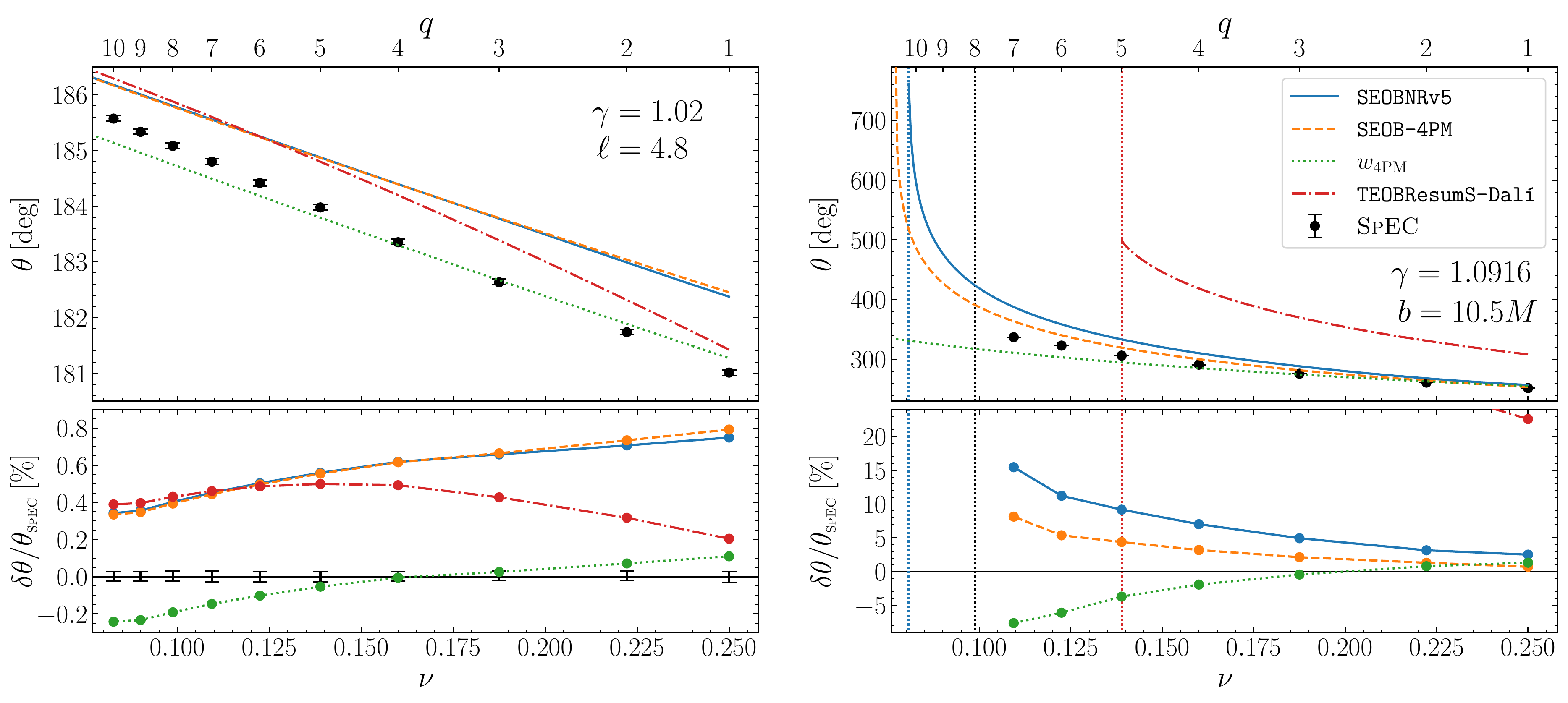}
  \caption{Comparison of EOB scattering angle results with those calculated by \spec{} for unequal-mass non-spinning systems. 
    The top panels show the relative scattering angle for constant $\gamma\!=\!1.02$, $\ell\!=\!4.8$ ({\em left}) and $\gamma\!=\!1.0916$, $b\! =\! 10.5M$ ({\em right}).
  The bottom panels show the relative difference in the relative scattering angle as compared to the NR values. The vertical dotted lines on the right plot show the first confirmed capture. 
  }
  \label{Fig:UnequalMass}
  \end{figure*} 

  The right panel of \Fig{Spin} shows the
  scattering angle for the equal-magnitude, anti-parallel spin simulations.
  The upper plot shows that increasing the spin magnitudes has little effect
  on the scattering angle, as previously observed in studies of similar 
  systems~\cite{Rettegno:2023ghr}. The lower panel shows that the agreement between the 
  EOB models does not vary greatly when altering the spin magnitudes. The
  closed-form models agree with the NR results to $\lesssim 5\%$, whereas the 
  \texttt{SEOBNRv5HM} and \Dali{} models have relative differences 
  to the NR of $\sim9\%$ and $\sim 55\%$ respectively.

  \Fig{AntiSpinDiff} shows the individual scattering angles $\theta_i$ for each BH for the equal magnitude, anti-parallel simulations.
  From the top panel, we see that the scattering angle across the range of spins is almost identical,
  only varying by fractions of a percent. Nevertheless, the difference in the scattering angles
  between the two BHs is apparent. The lower plot shows that the difference between the two angles grows
  as the magnitude of the spins increases the asymmetric emission in the system.
  Although previous NR results for unequal aligned spins exist~\cite{Rettegno:2023ghr,Albanesi:2024xus}, 
  this work is the first to directly measure the BHs' scattering angle difference, 
  which, being $< 0.1\%$ of the total angle, requires the high accuracy of the \spec{} results.
  The small magnitude of this difference is unsurprising, 
  as this effect emerges only at 5PM order and is therefore extremely subdominant~\cite{Buonanno:2024vkx}.

\subsection{Unequal-mass non-spinning black holes}

  In this section we consider unequal-mass non-spinning systems. We show results for two different
  slices of the parameter space: one with constant $\gamma = 1.02$ and $\ell=4.8$, 
  and another with constant $\gamma = 1.0916$ and $b = 10.5 M$. We fix $\ell$
  in one case and $b$ in the other in order to explore how each parameter behaves as we approach the
  geodesic limit.

  The left panel of \Fig{UnequalMass} presents the 
  scattering angle as a function of the symmetric mass ratio 
  for simulations with $\gamma = 1.02$ and $\ell=4.8$.
  The upper plot shows that while a decrease in the symmetric mass ratio does slightly increase the scattering angle,
  the change across the range of mass ratios is small. The lower panel shows that all of the EOB models
  capture this behavior well and are all within less than a percent of the NR results across all the mass ratios 
  with $w_{4{\rm PM}}$ showing the best agreement throughout.
  One interesting thing to note is that the \SEOBNR and \SEOBfourPM{} models tend towards each
  other as the mass ratio increases ($\nu$ decreases).
  This is due to the fact that both models are designed to recover the test mass limit.

  The right panel of \Fig{UnequalMass} shows the scattering angle for simulations with $\gamma = 1.0916$ and $b=10.5M$.
  Interestingly the behavior for these parameters is very different
  from the one just discussed.
  The upper plot shows that, as before, a decreasing symmetric mass ratio
  increases the scattering angle, but the change is very significant,
  including a capture at $q=8$ ($\nu \approx 0.1$). Both evolution models
  \SEOBNR{} and \Dali{} fail to accurately predict the capture with the former predicting
  a capture at $q \approx 10$ and the latter predicting a capture at $q \approx 5$.
  The lower panel shows that most of the EOB models
  are primarily within a few percent of the NR results at comparable masses but
  diverge as the symmetric mass ratio decreases. The noteable exception is the \Dali{} model which shows worse agreement
  across the whole range of mass ratios.
  These increased differences across all models are likely due to the fact that we 
  are approaching the separatrix and thus in the strong-field regime where we expect the accuracy of
  all the models to decrease.

\begin{figure*}
  \centering
  \includegraphics[width=0.97\linewidth,trim=0 10 0 12]{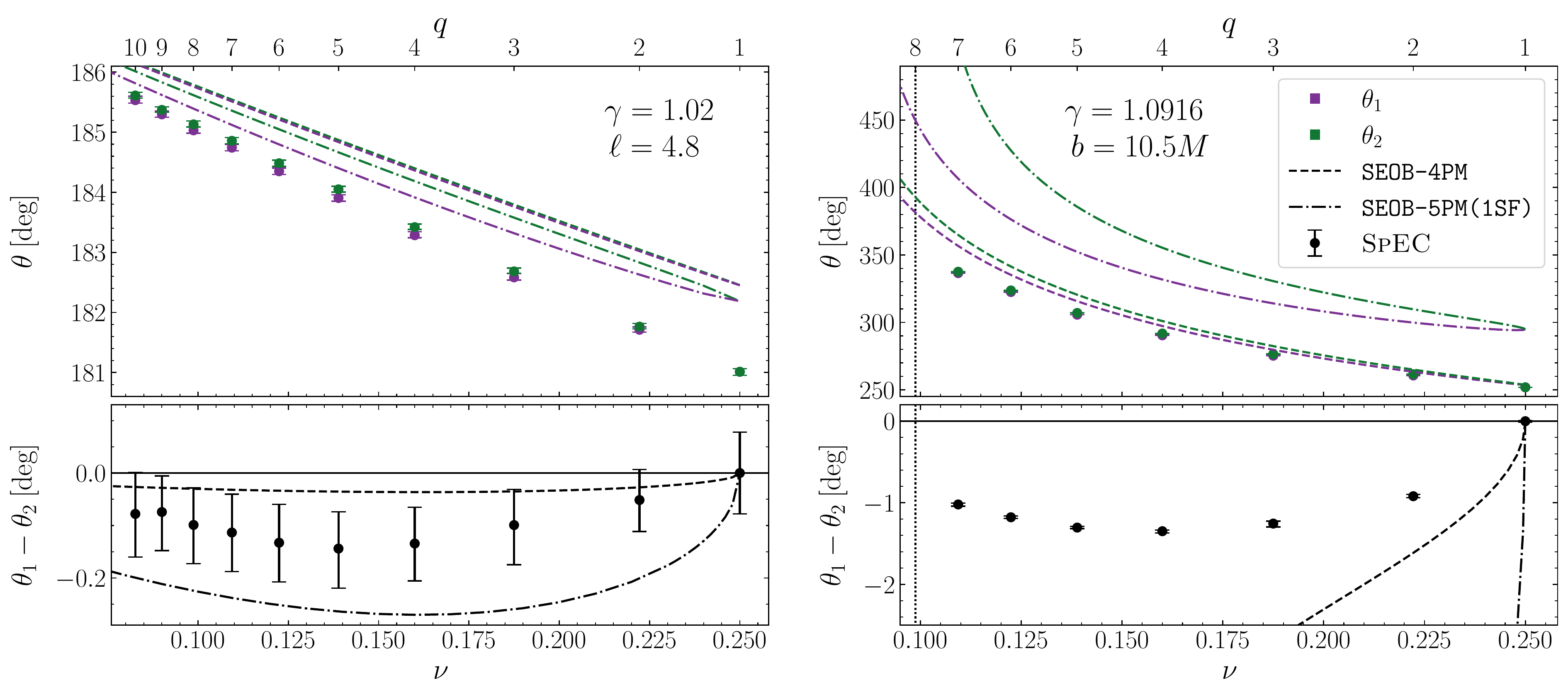}
  \caption{Scattering angle results for unequal mass non-spinning systems with constant 
  ($\gamma\!=\!1.02$, $\ell\!=\!4.8$) ({\em left}) and ($\gamma\!=\!1.0916$, $b \!=\! 10.5M$) ({\em right}).
  $\theta_1$ ($\theta_2$) represents the scattering angle of the larger (smaller) BH.
  The difference in the scattering angles is shown in the lower panel. 
  The vertical dotted line on the right plot shows the first confirmed capture of the \spec{} simulations. 
  In these comparisons, we use the asymmetrical closed-form \texttt{SEOB-PM} models based on the individual scattering angle $\theta_i$.
  }
  \label{Fig:UnequalMassDiff}
  \end{figure*} 

  The unequal-mass systems also break the symmetry of the system and lead to differences in the
  scattering angles of the two BHs. Figure~\ref{Fig:UnequalMassDiff} shows that for both unequal-mass systems the difference in the scattering angles increases when moving away from the comparable mass case due to the increased asymmetric emission in the systems.
  The difference in the angle peaks at $\nu \simeq 0.15$ and then decreases as the mass ratio becomes more extreme. One observation we can make is the size of the difference of the angles at this peak.
  The lower energy dataset ($\gamma = 1.02$) has a peak relative difference of $\sim 0.1\%$ while the higher energy dataset
  ($\gamma = 1.0916$) has a peak relative difference of $\sim 0.5\%$. This is likely due to the fact that the higher energy
  dataset is further into the strong field and thus has increased asymmetric emission.

Figure~\ref{Fig:UnequalMassDiff} also shows the estimates of each BH angle from the asymmetrical \SEOBPM{} model based on the individual scattering angles $\theta_i$.
  The lower energy dataset (left) shows that both models follow the same general shape of the NR data,
  with zero difference in the angles at equal mass, followed by the difference rising quickly to a peak and then tending towards equal
  angles in the test body limit. However, when comparing to the NR data, the 4PM model underestimates 
  the difference in the angles whereas the addition of the 5PM(1SF) terms causes the model to overestimate. 
  The higher energy dataset (right) shows very different behavior. Both models have the zero difference
  in the equal mass limit but then quickly diverge as the mass ratio increases. We hypothesize that
  this is due to the fact that we are approaching the separatrix, beyond which the notion of 
  two scattering angles does not make sense.

\section{Conclusions}
\label{sec:Conclusion}

In this work, we present the first set of unbound BBH simulations
generated using \spec{}. We explore several regions of parameter space including binaries with equal 
masses, aligned spin, and unequal masses. These new results 
provide the most accurate NR calculations of scattering angles in the literature, 
improving upon previous studies and extending the available 
parameter space.

We perform the first direct comparison of scattering angles between different 
NR codes, namely \spec{} and the \ETK{}. This comparison 
demonstrates good agreement between the two codes,
 strengthening confidence in the accuracy and reliability of unbound NR simulations.
 We encourage other groups within the  NR community to perform more direct comparisons 
 between different codes across a larger range of parameter space, and in different 
 physical observables, to further validate the results of future NR simulations. To aid this,
 we have recently made two scattering simulations (SXS:BBH:3999 and SXS:BBH:4292) 
 and one dynamical capture (SXS:BBH:4000) public as part of the SXS Collaboration's catalog of BBH simulations~\cite{SXSCatalogwebsite, Scheel:2025jct,SXS:BBH:3999,SXS:BBH:4292,SXS:BBH:4000}. 
We plan to release more unbound simulations alongside future \spec{} publications.

We compare our NR scattering angle results to a variety of EOB models, 
including the closed-form models \SEOBfourPM{} and $w_{\rm 4PM}$. We find that these
models generally perform well and agree with the NR results to within a 5\% deviation,
except near the scatter-capture separatrix. We also compare to the evolution models \SEOBNR{} and \Dali{}.
Our analysis shows that the \SEOBNR{} model has a similar level of accuracy to the closed-form models
in the non-spinning cases, while \Dali{} behaves significantly poorer 
in parts of the parameter space, especially at higher energies. 
This is somewhat surprising as \Dali{} includes eccentric corrections to the radiation-reaction fluxes,
which are expected to improve the accuracy of the EOB models relative to only using quasicircular fluxes, as in \SEOBNR{}.
The fidelity of both evolution models drops rapidly when introducing 
spins, with predictions of the scatter-capture separatrix occurring at 
significantly different spins than the NR results (see Fig.~\ref{Fig:Spin}).

 The key result of this study is the first measurement of disparate scattering
 angles from NR simulations due to the asymmetric emission of GWs. We have shown that 
 this effect, though small ($\lesssim1\%$ of the total angle), is extractable from \spec{} simulations for both anti-parallel spin systems and
  unequal mass systems. We also present the first closed-form models constructed to calculate the scattering angle from a single BH, rather than the relative angle of both BHs. Comparing to the NR results shows that these \SEOBPM{} models struggle to accurately predict the difference in the scattering angles due to it being a high-order perturbative effect.

  Having done the groundwork for NR scatter simulations from \spec{}, it would be an exciting future perspective to extend the parameter space.
  Here, a particularly intriguing direction would be scattering orbits in the high-energy limit.
  Importantly, in this limit, the PM regime is no longer valid and new theoretical approximations would be required to inform the EOB models.
  Namely, as pointed out by D'Eath~\cite{DEath:1976bbo} and then Kovacs and Thorne~\cite{Kovacs:1978eu}, perturbative scattering results are only valid provided
  that the overall deflection angle is small compared with the inverse boost parameter $\gamma^{-1}$ which in the high-energy limit implies:
  \begin{align}
      \frac{GE}{|b|}\ll\frac1\gamma\,.
  \end{align}
  With respect to the scattering angle, this requirement manifests itself in the form of a high-energy divergence as $\gamma\to\infty$,
  beginning at 4PM order.
  Thus, one would expect the closed-form models based on 4PM data and beyond to become more and more inaccurate in the high-energy regime which, indeed, was observed in Ref.~\cite{Swain:2024ngs}.

Looking ahead, one possible use of our results is the calibration of bound evolved models. 
The scattering angle, being a well-defined observable in the unbound regime, 
provides a clean constraint on the dynamics of the binary. By incorporating NR-calculated scattering 
angles into the calibration of the EOB Hamiltonian, we can refine the potential governing the motion 
of the bodies, ensuring more accurate predictions of the gravitational waveforms emitted by the binary.

\section*{Acknowledgements}

The authors would like to thank B.\ Leather, P.\ Lynch, and P.J.\ Nee for fruitful discussions.
G.M.~is supported by The Royal Society under grant URF\textbackslash R1\textbackslash 231578,
``Gravitational Waves from Worldline Quantum Field Theory''.
Computations were performed on the HPC system Urania at the Max Planck Computing and Data Facility.
H.R.R. acknowledges financial support provided
under the European Union’s H2020 ERC Advanced Grant ``Black holes:
gravitational engines of discovery'' grant agreement no.\@
Gravitas–101052587.  Views and opinions expressed are however those of
the authors only and do not necessarily reflect those of the European
Union or the European Research Council.  Neither the European Union
nor the granting authority can be held responsible for them.
A. R.-B. is supported by the Veni research programme which is
(partly) financed by the Dutch Research Council (NWO) under the grant
VI.Veni.222.396; acknowledges support from the Spanish Agencia Estatal
de Investigación grant PID2022-138626NB-I00 funded by
MICIU/AEI/10.13039/501100011033 and the ERDF/EU; is supported by the
Spanish Ministerio de Ciencia, Innovación y Universidades (Beatriz
Galindo, BG23/00056) and co-financed by UIB.
The figures in this article were produced with \textsc{matplotlib}~\cite{matplotlib1,matplotlib2}.

\appendix

\section{Effect of junk radiation}
\label{App:Junk}

One of the primary advantages of studying unbound systems is the 
ability to directly compare orbits that are unambiguously defined
using asymptotic quantities. However, this comparison is potentially
limited by the accuracy with which these asymptotic quantities are known.  In particular,
the junk radiation emitted during the initial relaxation of the
numerical simulations has the potential to introduce errors into
the scattering-angle comparisons. In this appendix, we investigate
the effect of junk radiation on the scattering angle calculations and
show that the effect is small and does not significantly impact the
results.

One disadvantage of the AMR settings discussed in Sec.~\ref{sec:AMR} is that
the junk radiation emitted during the initial relaxation of the numerical 
simulations is not resolved and thus cannot be accurately measured. Instead,
we gauge the effect of junk radiation by using the Newtonian approximations 
for the energy and angular momentum for non-spinning systems:
\begin{align}
  E_{\rm Newt} =& \: M + \frac{1}{2} \mu \: \vec{v} \cdot \vec{v} -\frac{\mu}{D}, \\
  \vec{J}_{\rm Newt} =& \: \mu \: \vec{D} \times \vec{v},
\end{align}
where $\vec{D}$ is the displacement vector between the two BHs, 
$D = |\vec{D}|$, 
and $\vec{v}$ is the relative velocity. We can use these formulae
to estimate the changes in the energy and the angular momentum of
two similar systems $A$ and $B$, which correspond to simulations 
at different initial separation $D_0$ targeting the identical 
physical configuration. It is useful
to split the difference in the velocities of 
each system into parallel and perpendicular components such that
\begin{equation}\label{eq:Deltav-decomp}
\Delta \vec{v} = \vec{v}_A - \vec{v}_B = \Delta v_{\parallel} \hat{v}_A + \Delta \vec{v}_{\perp},
\end{equation}
where $\hat{v}_A$ is the unit vector in the direction of $\vec{v}_A$ (since the relative velocities for the systems $A$ and $B$ are nearly identical, $\vec v_A\approx \vec v_B$, one can decompose in Eq.~(\ref{eq:Deltav-decomp}) also with respect to $\vec v_B$)x.
We can then express the change in the energy and angular momentum at fixed separation $D$ as
\begin{align}
  \Delta E =& \: \Delta M + \frac{1}{2} \mu^2 \left(\frac{\Delta m_1}{m_1^2} + \frac{\Delta m_2}{m_2^2}\right) \: \vec{v} \cdot \vec{v}+ \mu \: |\vec{v}| \: \Delta v_{\parallel}, \label{eq:DeltaE}\\
  |\Delta \vec{J}| =& \: \mu \left(\frac{\Delta m_1}{m_1^2} + \frac{\Delta m_2}{m_2^2}\right) |\vec{J}_{\rm ADM}| + \mu D |\Delta \vec{v}_{\perp}|,\label{eq:DeltaJ}
\end{align}
where $\Delta$ represents the difference between the two simulations.

\begin{figure}
  \centering
  \includegraphics[width=\linewidth,trim=6 10 0 0,clip=true]{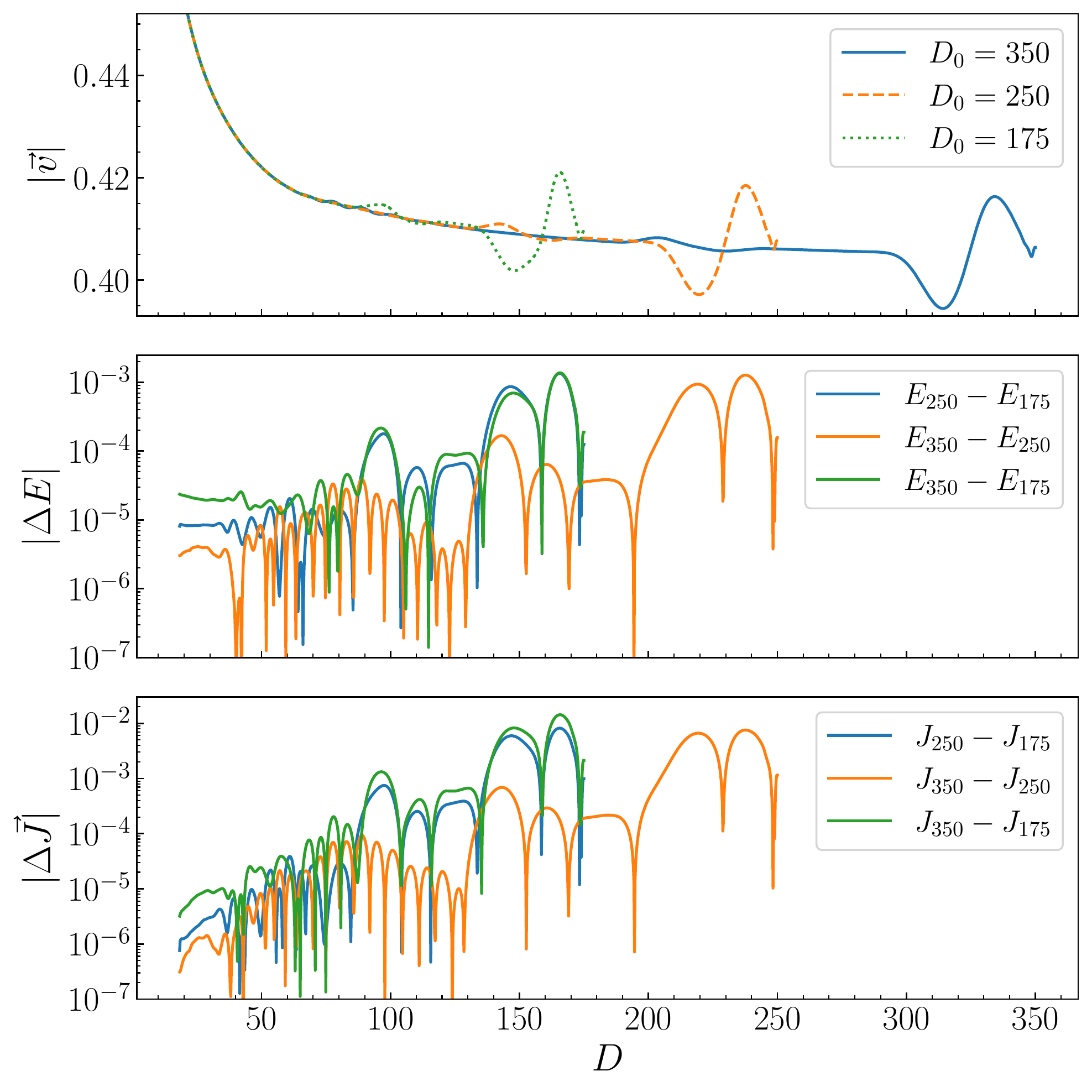}
  \caption{The differences between multiple runs
  with the same physical parameters but different initial separations.
  The top panel shows a zoom-in for the relative velocity of the two BHs
  as a function of separation on the ingoing leg. The bottom two panels
  show the difference in the energy and angular momentum of the simulations.
  }
  \label{Fig:Junk}
  \end{figure} 

In order to use these formulae to estimate the effect of the junk radiation, we
perform a series of trial runs with different initial separations 
and analyze the differences in the post-junk masses and trajectories. The results
of these simulations are shown in \Fig{Junk} for a set of simulations with
$(\Gamma, \ell, q) = (1.02264, 4.5367, 1)$ that start at initial separations
$D_0 = (175, 250, 350)$. The top panel of Fig.~\ref{Fig:Junk} shows the relative velocity of the two
BHs as a function of separation on the ingoing leg. Each curve shows an initial 
pulse at the start of the simulation as well as a secondary pulse when the junk 
radiation emitted from one BH interacts with the other. The middle and bottom panels
show the difference in the energy and angular momentum of the simulations 
from Eqs.~(\ref{eq:DeltaE}) and (\ref{eq:DeltaJ}). The curves show that the
differences between all of the configurations decay after the initial pulses 
to reach discrepancies of $\lesssim 10^{-5}$.

These results indicate that, despite the initial disturbances caused by junk 
radiation, the asymptotic parameters describing the orbit are accurate to
within approximately $10^{-5}$, and this is mirrored in Fig.~\ref{Fig:ETKComparison}. This provides 
confidence that the scattering angle calculations used in the analysis are 
robust against the effects of initial numerical transients, ensuring that 
comparisons between different simulations and models remain reliable.

\section{Details of the \texttt{SEOB-5PM(1SF)} model}
\label{App:5PMModel}

As discussed in the main text, the PM content of the closed-form models may be characterized by their accuracy after PM expanding both sides of Eq.~\eqref{eq:angleFromPR} which we reprint here for clarity:
\begin{align}\label{eq:matching}
    \theta+\pi
        =
        -2\int_{r_{\rm min}}^\infty\!{\rm d}r\frac{\partial}{\partial L}
        p_{r}
        \,.
\end{align}
The PM expansion of the right-hand-side is non-trivial but discussed e.g. in Refs.~\cite{Damour:1988mr,Damour:2019lcq,Buonanno:2024vkx}.
For the $n$PM closed-form models with $n\le4$ the PM expansion of the the right-hand-side of Eq.~\eqref{eq:matching} is simply required to match the analytic $n$PM angle of the left-hand-side.

As stated in the main text, the 5PM angle has a 2SF component which is not known.
Thus, in the PM matching of the \texttt{SEOB-5PM(1SF)} model, the $G^5$ part of the relative scattering angle is accurate only to 1SF order.
More precisely, using the notation of Ref.~\cite{Driesse:2024feo}, the analytic angle $\theta'_{\rm rel}$ used for matching is,
\begin{align}
  \theta'_{\rm rel}
\!&=\!
\Gamma
\sum_{n=1}^4
\bigg(\!\frac{GM}{b}\!\bigg)^n
\sum_{m=0}^{\lfloor\frac{n-1}{2}\rfloor}
\nu^m \theta^{(n,m)}
  +
\frac1{\Gamma}
\bigg(\!\frac{GM}{b}\!\bigg)^4
\nu^{2} \theta^{(4,2)}
\nonumber
\\
&+
\Gamma
\bigg(\!\frac{GM}{b}\!\bigg)^5
\Big[
  \theta^{(5,0)}
  +
  \nu \theta^{(5,1)}
  +
  \nu^2\frac{2+53 v^2}{5v^6}
\Big]
\ ,
\label{eq:5PMangle}
\end{align}
where $v=\sqrt{\gamma^2-1}/\gamma$ is the relative velocity.
The first line of Eq.~\eqref{eq:5PMangle} is the full 4PM relative angle while the next line gives the 1SF-accurate 5PM angle with additional 3PN-2SF terms.
We neglect spin because the asymmetric \texttt{SEOB-5PM(1SF)} model is used only for non-spinning comparisons, though spin effects would not add any further subtleties.
As discussed in the main text, the 3PN-2SF terms are included for self-consistency of the model.
Namely, those terms are determined from the 4PM input following the usual correspondence that $n$PM determines $(n-1)$PN.
If one would not include the 3PN-2SF terms in the matching angle $\theta'_{\rm rel}$, the 5PM(1SF) deformation would pick up unphysical poles in the $v\to0$ limit.

Finally, for the asymmetric \texttt{SEOB} models, we compute the individual scattering angle using the following relation,
\begin{align}\label{eq:rel1}
  \theta_1
  =
  \theta_{\rm rel}
  -
  \frac{E_1}{M\mu\sqrt{\gamma^2-1}}
      (\hat{\boldsymbol{b}}-\theta_{\rm rel}\hat{\boldsymbol{p}})
    \cdot
    \boldsymbol{P}_{\rm recoil}
    +\mathcal{O}(G^6)
  \ ,
\end{align}
which is similar to Eq.~\eqref{eq:averel} and uses the same notation.
An analogous equation for $\theta_2$ is obtained by symmetry.
For the \texttt{SEOB-5PM(1SF)} one requires the 5PM piece of the $\hat{\boldsymbol{b}}$-direction of the recoil which is 1SF-exact (see Ref.~\cite{Driesse:2024feo}).
The matching angle $\theta'_1$ for the asymmetric \texttt{SEOB-5PM(1SF)} model is then computed using Eq.~\eqref{eq:rel1} with $\theta_{\rm rel}$ given by $\theta_{\rm rel}'$ defined above and the recoil from Ref.~\cite{Driesse:2024feo} (throwing away any terms of order $G^6$ or beyond).

Finally, we note that the $G^5$ contributions were included in a PN-expanded manner.
Namely, they were expanded to $\mathcal{O}(v^{25})$, corresponding to 15PN order.
In the future it would be interesting to implement direct numerical evaluation of the multiple polylogarithms of maximal weight three required for these $G^5$ contributions.

\section{Scattering angle values}
\label{app:ScatteringAngles}

In this appendix, we provide the numerical values of the scattering 
angles for the datasets discussed in the main text. For symmetric systems, 
we give the average value of the scattering angles from the \spec{} simulations.
For the anti-parallel spin and unequal mass cases, we provide the angles for the individual BHs. All of the EOB values presented are for the relative scattering angle and
are calculated using the parameters of the \spec{} evolutions, as given in
the ancillary file \cite{data}.

\onecolumngrid

\begin{table}
  \renewcommand{\arraystretch}{1.2}
    \centering
    \begin{tabular}{c|c|cc|cc}
    \hline
\hspace{1.7em}$\ell$\hspace{1.7em} & \hspace{2.3em}\spec{}\hspace{2.3em} & \hspace{0.5em}\SEOBfourPM{}\hspace{0.5em} & \hspace{1em}$w_{\rm 4PM}$\hspace{1em} & \hspace{1em}\SEOBNR{}\hspace{1em} & \Dali{} \\ \hline
$4.3536$ & --- & $366.566$ & $379.778$ & $347.909$ & --- \\
$4.37640$ & $331.395^{+0.074}_{-0.145}$ & $333.730$ & $341.159$ & $326.648$ & --- \\
$4.39930$ & $306.963^{+0.017}_{-0.024}$ & $310.475$ & $315.322$ & $309.041$ & --- \\
$4.44510$ & $275.229^{+0.007}_{-0.008}$ & $277.857$ & $280.409$ & $281.053$ & $375.149$ \\
$4.53670$ & $236.282^{+0.006}_{-0.006}$ & $237.504$ & $238.547$ & $242.152$ & $277.114$ \\
$4.58260$ & $222.541^{+0.007}_{-0.006}$ & $223.390$ & $224.127$ & $227.772$ & $253.168$ \\
$4.67420$ & $201.202^{+0.006}_{-0.006}$ & $201.625$ & $202.036$ & $205.159$ & $220.508$ \\
$4.76590$ & $184.967^{+0.006}_{-0.006}$ & $185.178$ & $185.430$ & $187.909$ & $198.199$ \\
$5.04080$ & $152.153^{+0.007}_{-0.007}$ & $152.151$ & $152.232$ & $153.395$ & $157.516$ \\
$5.49910$ & $120.848^{+0.008}_{-0.008}$ & $120.806$ & $120.826$ & $121.210$ & $122.425$ \\
$5.95730$ & $101.726^{+0.008}_{-0.008}$ & $101.693$ & $101.700$ & $101.863$ & $102.216$ \\
$6.41560$ & $88.443^{+0.009}_{-0.009}$ & $88.420$ & $88.423$ & $88.509$ & $88.567$ \\
$6.87390$ & $78.532^{+0.010}_{-0.010}$ & $78.515$ & $78.517$ & $78.570$ & $78.527$ \\
$7.33210$ & $70.787^{+0.011}_{-0.010}$ & $70.775$ & $70.775$ & $70.812$ & $70.741$ \\ \hline
    \end{tabular}
    \caption{Scattering angles (in degrees) for equal mass, non-spinning BHs with $\Gamma = 1.02264$. A dash (---) indicates a capture.}
\end{table}

\begin{table}
  \renewcommand{\arraystretch}{1.2}
    \centering
    \begin{tabular}{c|c|cc|cc}
    \hline
\hspace{1.7em}$\Gamma$\hspace{1.7em} & \hspace{2.3em}\spec{}\hspace{2.3em} & \hspace{0.5em}\SEOBfourPM{}\hspace{0.5em} & \hspace{1em}$w_{\rm 4PM}$\hspace{1em} & \hspace{1em}\SEOBNR{}\hspace{1em} & \Dali{} \\ \hline
$1.00250$ & $216.341^{+0.189}_{-0.233}$ & $220.597$ & $217.009$ & $219.492$ & $216.167$ \\
$1.00500$ & $200.857^{+0.051}_{-0.062}$ & $203.263$ & $200.843$ & $202.537$ & $201.507$ \\
$1.00750$ & $194.116^{+0.025}_{-0.032}$ & $196.264$ & $194.314$ & $195.941$ & $196.409$ \\
$1.01000$ & $192.194^{+0.017}_{-0.021}$ & $193.981$ & $192.344$ & $194.060$ & $196.123$ \\
$1.01250$ & $193.094^{+0.012}_{-0.015}$ & $194.587$ & $193.235$ & $195.137$ & $199.136$ \\
$1.01500$ & $196.034^{+0.010}_{-0.012}$ & $197.319$ & $196.280$ & $198.463$ & $204.940$ \\
$1.01750$ & $200.799^{+0.008}_{-0.010}$ & $201.848$ & $201.197$ & $203.751$ & $213.545$ \\
$1.02000$ & $207.177^{+0.007}_{-0.008}$ & $208.084$ & $207.945$ & $210.955$ & $225.383$ \\
$1.02250$ & $215.426^{+0.007}_{-0.006}$ & $216.130$ & $216.701$ & $220.217$ & $241.505$ \\
$1.02500$ & $225.684^{+0.006}_{-0.005}$ & $226.259$ & $227.856$ & $231.846$ & $264.086$ \\
$1.02750$ & $238.410^{+0.005}_{-0.004}$ & $239.025$ & $242.190$ & $246.418$ & $298.436$ \\
$1.03000$ & $254.569^{+0.005}_{-0.004}$ & $255.415$ & $261.165$ & $264.894$ & $363.341$ \\
$1.03250$ & $275.734^{+0.004}_{-0.004}$ & $277.353$ & $287.933$ & $289.013$ & --- \\
$1.03500$ & $305.705^{+0.006}_{-0.006}$ & $309.117$ & $331.114$ & $322.238$ & --- \\
$1.03750$ & $358.221^{+0.056}_{-0.122}$ & $363.898$ & $438.527$ & --- & --- \\
$1.0385$ & --- & $403.006$ & $813.250$ & --- & --- \\ \hline
    \end{tabular}
    \caption{Scattering angles (in degrees) for equal mass, non-spinning BHs with $\ell = 4.608$. A dash (---) indicates a capture.}
\end{table}

\begin{table}
  \renewcommand{\arraystretch}{1.2}
    \centering
    \begin{tabular}{cc|c|cc|cc}
    \hline
\hspace{1.2em}$\chi_1$\hspace{1.2em} & \hspace{1.2em}$\chi_2$\hspace{1.2em} & \hspace{2.3em}\spec{}\hspace{2.3em} & \hspace{0.5em}\SEOBfourPM{}\hspace{0.5em} & \hspace{1em}$w_{\rm 4PM}$\hspace{1em} & \hspace{1em}\SEOBNR{}\hspace{1em} & \Dali{} \\ \hline
$-0.26$ & $-0.26$  & --- & $405.019$ & $510.560$ & --- & --- \\
$-0.25$ & $-0.25$ & $385.441^{+0.020}_{-0.047}$ & $380.369$ & $437.515$ & --- & --- \\
$-0.20$ & $-0.20$ & $300.868^{+0.009}_{-0.005}$ & $310.446$ & $323.630$ & --- & --- \\
$-0.10$ & $-0.10$ & $238.782^{+0.007}_{-0.005}$ & $247.779$ & $251.214$ & --- & --- \\
$\phantom{-}0.00$ & $\phantom{-}0.00$ & $207.041^{+0.005}_{-0.004}$ & $214.149$ & $216.177$ & $226.521$ & $319.800$ \\
$\phantom{-}0.10$ & $\phantom{-}0.10$ & $186.111^{+0.005}_{-0.005}$ & $191.701$ & $193.518$ & $172.404$ & $238.814$ \\
$\phantom{-}0.20$ & $\phantom{-}0.20$ & $170.712^{+0.007}_{-0.008}$ & $175.156$ & $177.033$ & $151.030$ & $203.180$ \\
$\phantom{-}0.30$ & $\phantom{-}0.30$ & $158.656^{+0.009}_{-0.011}$ & $162.224$ & $164.215$ & $137.689$ & $181.149$ \\
$\phantom{-}0.40$ & $\phantom{-}0.40$ & $148.875^{+0.010}_{-0.013}$ & $151.717$ & $153.814$ & $127.987$ & $165.616$ \\
$\phantom{-}0.50$ & $\phantom{-}0.50$ & $140.713^{+0.012}_{-0.016}$ & $142.941$ & $145.113$ & $120.399$ & $153.841$ \\
$\phantom{-}0.60$ & $\phantom{-}0.60$ & $133.738^{+0.014}_{-0.019}$ & $135.508$ & $137.725$ & $113.860$ & $144.609$ \\ \hline
    \end{tabular}
    \caption{Scattering angles (in degrees) for equal mass BHs with spin aligned or anti-aligned with the orbital angular momentum. All of these simulations have constant $\Gamma = 1.055$ and $\ell = 5.18$. A dash (---) indicates a capture.}
\end{table}

\begin{table}
  \renewcommand{\arraystretch}{1.2}
    \centering
    \begin{tabular}{cc|cc|cc|cc}
    \hline
\hspace{1.2em}$\chi_1$\hspace{1.2em} & \hspace{1.2em}$\chi_2$\hspace{1.2em} & \hspace{1.3em}\spec{} ($\theta_1$)\hspace{1.3em} & \hspace{1.3em}\spec{} ($\theta_2$)\hspace{1.3em} & \hspace{0.5em}\SEOBfourPM{}\hspace{0.5em} & \hspace{1em}$w_{\rm 4PM}$\hspace{1em} & \hspace{1em}\SEOBNR{}\hspace{1em} & \Dali{} \\ \hline
$\phantom{-}0.00$ & $0.00$ & $207.039^{+0.005}_{-0.004}$ & $207.044^{+0.005}_{-0.004}$ & $214.149$ & $216.177$ & $226.521$ & $319.800$ \\
$-0.10$ & $0.10$ & $207.044^{+0.007}_{-0.004}$ & $207.018^{+0.006}_{-0.006}$ & $214.085$ & $216.142$ & $226.463$ & $319.673$ \\
$-0.20$ & $0.20$ & $207.026^{+0.011}_{-0.006}$ & $206.974^{+0.009}_{-0.010}$ & $213.915$ & $216.056$ & $226.315$ & $319.383$ \\
$-0.30$ & $0.30$ & $206.995^{+0.015}_{-0.009}$ & $206.918^{+0.011}_{-0.013}$ & $213.635$ & $215.918$ & $226.073$ & $318.918$ \\
$-0.40$ & $0.40$ & $206.940^{+0.020}_{-0.012}$ & $206.836^{+0.013}_{-0.015}$ & $213.256$ & $215.735$ & $225.746$ & $318.317$ \\
$-0.50$ & $0.50$ & $206.828^{+0.026}_{-0.015}$ & $206.699^{+0.015}_{-0.018}$ & $212.737$ & $215.467$ & $225.288$ & $317.357$ \\ \hline
    \end{tabular}
    \caption{Scattering angles (in degrees) for equal mass BHs with spin parallel to the orbital angular momentum. All of these simulations have constant $\Gamma = 1.055$ and $\ell = 5.18$.}
\end{table}

\begin{table}
  \renewcommand{\arraystretch}{1.2}
    \centering
    \begin{tabular}{c|cc|cc|cc}
    \hline
\hspace{1em}$q$\hspace{1em} & \hspace{1.3em}\spec{} ($\theta_1$)\hspace{1.3em} & \hspace{1.3em}\spec{} ($\theta_2$)\hspace{1.3em} & \hspace{0.5em}\SEOBfourPM{}\hspace{0.5em} & \hspace{1em}$w_{\rm 4PM}$\hspace{1em} & \hspace{1em}\SEOBNR{}\hspace{1em} & \Dali{} \\ \hline
$1.0$ & $181.013^{+0.050}_{-0.060}$ & $181.013^{+0.050}_{-0.060}$ & $182.447$ & $181.211$ & $182.369$ & $181.383$ \\
$2.0$ & $181.712^{+0.045}_{-0.040}$ & $181.764^{+0.054}_{-0.037}$ & $183.046$ & $181.841$ & $182.996$ & $182.288$ \\
$3.0$ & $182.584^{+0.061}_{-0.044}$ & $182.683^{+0.057}_{-0.029}$ & $183.797$ & $182.630$ & $183.786$ & $183.364$ \\
$4.0$ & $183.285^{+0.057}_{-0.042}$ & $183.419^{+0.052}_{-0.039}$ & $184.415$ & $183.275$ & $184.418$ & $184.188$ \\
$5.0$ & $183.905^{+0.052}_{-0.054}$ & $184.049^{+0.049}_{-0.047}$ & $184.925$ & $183.806$ & $184.934$ & $184.823$ \\
$6.0$ & $184.350^{+0.052}_{-0.053}$ & $184.482^{+0.050}_{-0.051}$ & $185.265$ & $184.161$ & $185.278$ & $185.245$ \\
$7.0$ & $184.742^{+0.054}_{-0.051}$ & $184.856^{+0.055}_{-0.049}$ & $185.563$ & $184.471$ & $185.578$ & $185.594$ \\
$8.0$ & $185.031^{+0.055}_{-0.049}$ & $185.130^{+0.056}_{-0.044}$ & $185.759$ & $184.676$ & $185.775$ & $185.826$ \\
$9.0$ & $185.296^{+0.055}_{-0.049}$ & $185.371^{+0.050}_{-0.040}$ & $185.940$ & $184.863$ & $185.956$ & $186.030$ \\
$10.0$ & $185.532^{+0.066}_{-0.050}$ & $185.610^{+0.052}_{-0.044}$ & $186.153$ & $185.083$ & $186.167$ & $186.254$ \\ \hline
    \end{tabular}
    \caption{Scattering angles (in degrees) for unequal mass, non-spinning BHs with $\gamma\!=\!1.02$ and $\ell\!=\!4.8$. \spec{} values are given for both BHs, with $\theta_1$ corresponding to the larger mass. EOB values correspond to the relative scattering angle.}
\end{table}

\begin{table}
  \renewcommand{\arraystretch}{1.2}
    \centering
    \begin{tabular}{c|cc|cc|cc}
    \hline
\hspace{1em}$q$\hspace{1em} & \hspace{1.3em}\spec{} ($\theta_1$)\hspace{1.3em} & \hspace{1.3em}\spec{} ($\theta_2$)\hspace{1.3em} & \hspace{0.5em}\SEOBfourPM{}\hspace{0.5em} & \hspace{1em}$w_{\rm 4PM}$\hspace{1em} & \hspace{1em}\SEOBNR{}\hspace{1em} & \Dali{} \\ \hline
$1.0$ & $251.864^{+0.006}_{-0.007}$ & $251.864^{+0.006}_{-0.007}$ & $253.658$ & $255.176$ & $258.186$ & $308.768$ \\
$2.0$ & $260.639^{+0.020}_{-0.015}$ & $261.558^{+0.009}_{-0.011}$ & $264.510$ & $263.184$ & $269.326$ & $331.792$ \\
$3.0$ & $275.300^{+0.028}_{-0.041}$ & $276.555^{+0.009}_{-0.013}$ & $281.810$ & $274.723$ & $289.566$ & $370.643$ \\
$4.0$ & $290.409^{+0.010}_{-0.019}$ & $291.757^{+0.010}_{-0.012}$ & $300.381$ & $285.487$ & $311.518$ & $419.112$ \\
$5.0$ & $305.691^{+0.010}_{-0.004}$ & $306.995^{+0.012}_{-0.013}$ & $319.748$ & $295.001$ & $334.444$ & $497.047$ \\
$6.0$ & $322.503^{+0.012}_{-0.005}$ & $323.680^{+0.013}_{-0.012}$ & $340.448$ & $303.412$ & $359.369$ & --- \\
$7.0$ & $336.520^{+0.011}_{-0.007}$ & $337.541^{+0.019}_{-0.012}$ & $364.464$ & $311.312$ & $389.126$ & --- \\
$8.0$ & --- & --- & $392.833$ & $318.261$ & $425.986$ & --- \\ \hline\hline
    \end{tabular}
    \caption{Scattering angles (in degrees) for unequal mass, non-spinning BHs with $\gamma\!=\!1.0916$ and $b\! =\! 10.5M$. \spec{} values are given for both BHs, with $\theta_1$ corresponding to the larger mass. EOB values correspond to the relative scattering angle. A dash (---) indicates a capture.}
\end{table}

\clearpage
\twocolumngrid
\bibliography{biblio}

\end{document}